\documentclass[journal,twocolumn]{IEEEtran}

\usepackage{cite}
\usepackage{amsmath,amssymb,amsfonts}
\usepackage{algorithmic}
\usepackage{graphicx}
\usepackage{subcaption}
\usepackage{longtable}
\usepackage{amsmath}

\begin{document}

\newtheorem{thm}{Theorem}
\newtheorem{lem}{Lemma}
\newtheorem{cor}{Corollary}
\newtheorem{defn}{Definition}

\title{Polarization Reconfigurable Transmit-Receive Beam Alignment with Interpretable Transformer}
\author{\IEEEauthorblockN{Seungcheol Oh{$^\dagger$$^\ddagger$}, Han Han, Joongheon Kim$^\ddagger$, Sean Kwon{$^\dagger$}} \\
{$^\dagger$}Department of Electrical Engineering, California State University - Long Beach\\
{$^\ddagger$}School of Electrical Engineering, Korea University\\
Marvell Semiconductor Canada, Inc.\\
Email: seungoh@korea.ac.kr,{~}  hanhan@marvell.com, {~}joongheon@korea.ac.kr, {~}sean.kwon@csulb.edu
\thanks{Part of this work is supported by South Korean Ministry of Trade, Industry and Energy (RS-2024-00444136). Sean Kwon and Joongheon Kim are the corresponding authors.}
}
\maketitle

\vspace{-1cm}
\begin{abstract}
Recent advancement in next generation reconfigurable antenna and fluid antenna technology has influenced the wireless system with polarization reconfigurable (PR) channels to attract significant attention for promoting beneficial channel condition. We exploit the benefit of PR antennas by integrating such technology into massive multiple-input-multiple-output (MIMO) system. In particular, we aim to jointly design the polarization and beamforming vectors on both transceivers for simultaneous channel reconfiguration and beam alignment, which remarkably enhance the beamforming gain. However, joint optimization over polarization and beamforming vectors  without channel state information (CSI) is a challenging task, since depolarization increases the channel dimension; whereas massive MIMO systems typically have low-dimensional pilot measurement from limited radio frequency (RF) chain.
This leads to pilot overhead because the transceivers can only observe low-dimensional measurement of the high-dimension channel. This paper pursues the reduction of the pilot overhead in such systems by proposing to employ \emph{interpretable transformer}-based deep learning framework on both transceivers to actively design the polarization and beamforming vectors for pilot stage and transmission stage based on the sequence of accumulated received pilots. 
Numerical experiments demonstrate the significant performance gain of our proposed framework over the existing non-adaptive and active data-driven methods. 
Furthermore, 
we exploit the interpretability of our proposed framework to analyze the learning capabilities of the model. 
\end{abstract}

\begin{IEEEkeywords}
Beamforming, polarization reconfiguration, transformer, machine learning, pilot, multiple-input-multiple-output (MIMO)
\end{IEEEkeywords}

\section{Introduction}
Next generation reconfigurable antenna technologies have fostered a new era of wireless communication systems consisting with novel schemes and inherent new paradigm. Fluid antennas are regarded as one of the most representative next generation reconfigurable antennas, which can tailor the radio propagation characteristics, \textit{inter alia}, rotation of the antenna orientation to create diverse polarization \cite{new2024tutorial}. In designing wireless systems, utilizing the polarization domain has gained significant attention, and it can be supported by polarization reconfigurable (PR) antennas with novel materials, such as graphene metasurfaces and liquid metals \cite{new2024tutorial,PRimplementation1,RPimplementation2}.

Fluid antennas, in particular, liquid metal-based PR antennas introduced in \cite{new2024tutorial,PRimplementation1,RPimplementation2},  offer a more compact design and better operational flexibility than switch-based PR antennas reported in \cite{bulkyswitch-antenna1, bulkswitch-Pol2}. Hence, the integration of the former into existing wireless systems is more feasible and attractive than the latter. This has opened up possibilities to fully leverage the polarization domain in the design of wireless communication systems, providing additional degrees of freedom for enhanced multiplexing and diversity gain. Consequently, numerous studies have investigated the benefit of exploiting polarization domain \cite{Kwon_Molisch_Globecom,AntennaSelection,Heath_LinearPol, Kwon_PR_NOMA_Asilomar24, Oh_Kwon_MPS_Journal, gutierrez2020novelPSM, Kwon_Stuber_PDMA, Kwon_Stuber_GeoTheory, Kwon_PR_NOMA_Asilomar24, zhang2019Polusage, towards6G_MIMO,EmilDualPol}.

PR antennas provide a promising technology for next generation wireless communications, due to their capability of reconfiguring wireless channels between transceivers to promote beneficial channel state. This, in turn, effectively improves the wireless channel capacity \cite{Review_RA_2020, RA_Analysis_2022, Kwon_Molisch_Globecom, AntennaSelection, Kwon_PR_NOMA_Asilomar24, Heath_LinearPol, Oh_Kwon_MPS_Journal} or symbol error rate (SER)  \cite{Oh_Kwon_MPS_Journal, gutierrez2020novelPSM, Kwon_Stuber_PDMA}. Studies have recently been conducted to design polarization vectors for PR antennas applied to multi-antenna wireless communication systems. In particular, joint polarization pre/post-coding is proposed and utilized to significantly increase multiple-input-multiple-output (MIMO) channel capacity with optimal antenna polarization vectors in \cite{Kwon_Molisch_Globecom, AntennaSelection}; thereafter, the scheme is exploited in recent papers \cite{Kwon_PR_NOMA_Asilomar24, Heath_LinearPol}.

PR antennas are expected to be exploited in future wireless communication systems which consist of a large number of antenna elements, e.g., extremely large-scale MIMO (XL-MIMO) \cite{XL_MIMO, towards6G_MIMO}. Establishing reliable communication in such scenarios requires the joint design of polarization and beamforming vectors at both the transmitter (Tx) and receiver (Rx). However, this design heavily depends on accurate knowledge of channel state information (CSI) \cite{Kwon_Stuber_GeoTheory, Double_Direction}. In practice, obtaining the CSI in multi-PR-antenna systems is challenging, since PR wireless channel is typically higher-dimensional than the dimension of measured pilots. Moreover, large-scale multi-antenna systems often operate with a limited number of radio frequency (RF) chains due to hardware constraints \cite{molisch2004mimo, AntennaSelection}, further complicating CSI acquisition by reducing the dimensionality of the measured pilots.

This paper takes into account a joint polarization reconfiguration and beam alignment (PR-BA) problem in a point-to-point massive PR-MIMO channel and system. In this setting, multi-PR-antenna transceivers jointly optimize their polarization and beamforming vectors based on received pilot signals to achieve polarization channel reconfiguration and beam alignment simultaneously, without the prior knowledge of the perfect CSI. Because the both ends are equipped with low number of RF chain, the design on both transceiver ends is based on limited number of low-dimensional pilot measurements. Therefore, the joint PR-BA problem addresses the limitation of the polarization design method proposed in \cite{Kwon_Molisch_Globecom, Heath_LinearPol}, which rely on the assumption of perfect CSI. 

The joint PR-BA problem is challenging to solve because it involves the estimation of the PR channel which is inherently higher-dimensional than what is directly observed through pilots at both transceiver ends. This is because PR systems exploit the PR antennas based on the channel depolarization matrix \cite{Kwon_Stuber_GeoTheory}; which comes with the cost of increase in the dimensionality of the channel compared to what can be observed through pilots. Furthermore, limited number of RF chains on both ends limits the observed pilot dimension. With such increase in the channel dimension with depolarization effect and reduction of observed pilot dimension with limited number of RF chain, channel estimation becomes highly non-trivial at both transceiver sides. This leads to excessive pilot training overhead as the scale of the multi-PR-antenna systems grow. 

The main idea of this paper is that significant savings on pilot overhead is possible by leveraging a deep learning framework. This proposed framework employs a ping-pong pilot scheme, where the Tx and Rx take turns sending pilots while simultaneously updating and utilizing their polarization and beamforming vectors based on the accumulated sequence of received pilots. The nature of ping-pong pilot scheme is sequential; thus, we propose to employ transformer architecture to take the advantage of the multi-head attention mechanism \cite{blockrecurrentTransformer, vaswani2017attention}. Specifically, we employ two separate transformer based deep learning units at the Tx and Rx to efficiently summarize and extract the temporally correlated information from the received pilots over multiple pilot stages to design the Tx and Rx polarization and beamforming vectors for both the pilot stage and the final data transmission stage. Simulation results show that the proposed transformer-based framework significantly reduces the pilot overhead compared to other existing data-driven methods that are based on deep neural network (DNN) and recurrent neural network (RNN). 

\subsection{Prior Work} 
Joint polarization channel reconfiguration and MIMO pre-post coding problem has been investigated in  \cite{Kwon_Molisch_Globecom, AntennaSelection, Heath_LinearPol}. The research in \cite{Kwon_Molisch_Globecom, AntennaSelection} propose an iterative polarization optimization (IPO) scheme for joint Tx-Rx polarization pre-post coding based on the theoretical closed-form solution at one end of Tx or Rx. The scheme in \cite{Kwon_Molisch_Globecom, AntennaSelection} significantly increases sum-rate by exploiting the inherent characteristics of wireless channel depolarization. The scheme in \cite{Kwon_Molisch_Globecom, AntennaSelection} is also fully utilized for the wideband system in \cite{Heath_LinearPol}. On the other hand, the design methods in \cite{Kwon_Molisch_Globecom,AntennaSelection, Heath_LinearPol} require perfect knowledge of the CSI. While \cite{Kwon_Molisch_Globecom, Heath_LinearPol} briefly suggest a simple pilot scheme which utilizes dual-polarization configuration for channel estimation, multi-PR-antenna system design under imperfect CSI is not investigated.

Multi-antenna system design that employs polarization diversity under imperfect CSI has been discussed in \cite{EmilDualPol, DNN_PMISO}. For instance, the joint design over the channel estimation and beamforming in a massive polarized-MIMO system is presented in \cite{EmilDualPol}. However, they only consider dual-polarized antennas in their system. On the other hand, \cite{DNN_PMISO} employs PR antenna elements to massive multiple-input-single-output (MISO) system and design the polarization and beamforming vectors based on the channel estimated using the least square (LS). While solution for the multi-PR-antenna system design under imperfect CSI is provided, their LS-based channel estimation suffers from the pilot training overhead. This is because the PR antenna elements increase the wireless channel dimensionality \cite{Kwon_Stuber_GeoTheory} which makes the pilot overhead more severe. To mitigate this, \cite{DNN_PMISO} employs deep neural networks (DNNs) on both transceiver ends to map the pilot signals to optimal polarization and beamforming vectors, significantly reducing the overhead compared to the LS-based method. However, their model assumes a fully digital Tx in a massive MISO setting, where majority of the dimension over the channel can be observed from the received pilots. This is impractical since a more practical setting would involve a limited number of RF chains for large-scale multi-antenna systems \cite{massiveMIMO, AntennaSelection}.

In this paper we aim to mitigate the pilot training overhead in multi-PR-antenna systems under imperfect CSI with a limited number of RF chains. Without considering polarization diversity, reducing the pilot training overhead in conventional large-scale multi-antenna wireless communication systems with a limited number of RF chains has been an active area of research \cite{BeamAlignment1, BeamAlignment2, BeamAlignment3, pingpong_2nd, pingpong_3rd}. A key approach to minimizing the pilot overhead is to design efficient beam alignment algorithms \cite{beamalignment0}, which estimate the angle of arrival (AoA) and the angle of departure (AoD) during the pilot stage, to align the beamforming vectors accordingly to achieve higher signal to noise ratio (SNR). For instance, \cite{BeamAlignment1, BeamAlignment2} propose beam alignment algorithms based on compressed sensing \cite{Compressed_Sensing}. Although effective, these algorithms are limited in performance due to their non-adaptive nature. Specifically, their beamforming vectors are randomly placed initially and fixed during the pilot stages, which can result in reduced SNR due to the lack of adaptation to the channel during the pilot stage.

Another approach is actively designing the beamforming vectors during the pilot stage \cite{BeamAlignment3, pingpong_1st, pingpong_2nd, pingpong_3rd}. Specifically, \cite{pingpong_1st, pingpong_2nd, pingpong_3rd} propose to utilize a ping-pong scheme during pilot stage where transceivers send pilots back and forth while actively designing the beamforming vectors based the accumulated received pilots. The ping-pong scheme based active beam aligning design demonstrates that it has significant advantage over the non-adaptive methods in further reducing the pilot overhead \cite{pingpong_1st, pingpong_2nd, pingpong_3rd}. 
However, such a ping-pong strategy become computationally intractable in the scenarios with multi-path channel, as the solution usually require calculating the posterior distribution of the AoA and AoD. This complexity motivates the need for alternative approaches that can efficiently handle the problem. 

One promising direction is to leverage machine learning techniques, which have been increasingly used to tackle computationally intractable problems in wireless communications \cite{MLApproach, MLbeamAlignment1,MLbeamAlignment2, Alkhateeb_GRU2,LSTM_ComputerVIsion_BeamAlignment}. Specifically, we are interested in machine learning solutions for sequential problem because ping-pong strategy actively designs the beamforming vectors based on the sequence of previously received pilot, which makes the problem sequential in nature. 
The sequential problems that arise in wireless communication systems have predominantly been addressed using RNN architectures, i.e., gated recurrent unit (GRU) and long short-term memory (LSTM) cell \cite{Alkhateeb_GRU1, Alkhateeb_GRU2,LSTM_ComputerVIsion_BeamAlignment}. For instance, GRU was employed for beam tracking and beam prediction 
\cite{Alkhateeb_GRU1, Alkhateeb_GRU2}, while LSTM was utilized for beam alignment \cite{LSTM_ComputerVIsion_BeamAlignment}. 

However, in recent years, transformers have demonstrated a superior performance over RNN-based framework in handling sequential tasks, particularly in natural language processing \cite{karita2019comparative}. Furthermore, \cite{Decision_Transformer} highlights the effectiveness of transformers in sequential decision making tasks. The performance gain of transformer over RNN is primarily attributed to \emph{multi-head attention} mechanism. With multi-head attention, the transformer extracts a more diverse contextual dependencies of the sequential data over RNN architectures \cite{vaswani2017attention}. This is an important architectural feature that motivates us to leverage transformer-based architecture to solve the joint PR-BA problem for this paper.


\subsection{Contributions}
The primary contributions of this paper is  summarized as follows. 
\subsubsection{Solution for joint PR-BA} we solve the joint PR-BA problem which jointly designs the polarization and beamforming vectors to simultaneously achieve polarization channel reconfiguration and beam alignment to maximize the beamforming gain under imperfect CSI. By solving this problem, we formally address the massive PR-MIMO system design with limited RF chain under imperfect CSI. The main problem of concern is pilot overhead which is caused by the increase in channel dimension from channel depolarization and reduction in observed pilot dimension from limited RF chain. We significantly overcome the overhead by employing a ping-pong scheme in which the Tx and Rx take turns sending pilot symbols every pilot stage while simultaneously utilizing and updating their polarization and beamforming vectors based on previously received pilot signals. This strategy achieves significant performance gain over benchmarks that utilizes non-adaptive strategy \cite{DNN_PMISO}.


\subsubsection{Transformer-based Solution}
This paper proposes a transformer-based framework to solve the joint PR-BA problem. Unlike RNN-based architectures, which sequentially update temporally correlated data in a hidden state vector, we utilize the \emph{multi-head attention} mechanism. This allows the model to extract diverse contextual dependencies in each attention head, enabling a richer understanding of temporally correlated pilot signals. We show the performance gain that our proposed transformer-based solution has over RNN solution in our evaluation. Furthermore, we show that the solution is generalizable over more complex channel with increased number of paths in multi-path channel environment. 

We mention here that deep learning approach to solve problems formulated with ping-pong pilot scheme under mmWave MIMO communication system has been previously presented in \cite{ping_pong}. However, they employ LSTM to capture the temporal correlation between the pilot signals, which performs significantly inferior to transformer. Furthermore, in this paper, we optimize polarization vectors in conjunction with beamforming vectors to jointly reconfigure the channel for beneficial wireless environment. We compare the performance in our experiment and demonstrate that transformers have advantage over RNN-based method.

\subsubsection{Interpretation}
In this paper, we provide insights into the learning capabilities of the transformer-based model, focusing on the attention scores and array response. We first analyze the impact of individual attention heads on the transformer's overall performance and compare these findings with those of a GRU-based RNN. Next, we examine the array responses generated by both models, offering an explanation for the transformer's superior performance. Specifically, we analyze the attention scores from each head in the multi-head attention mechanism and provide a detailed interpretation of the results.


\subsection{Organization of the Paper and Notations}
The remaining parts of this paper are organized as follows. Sec. \ref{sec:systemModelNProblemFormulation} introduces the system model, ping-pong pilot protocol and formulates the problem of interest. Sec. \ref{sec:RTF_solution} presents the proposed transformer-based framework to solve the joint PR-BA problem. Sec. \ref{sec:Numerical Experiments} and Sec. \ref{sec:InterpretationandDiscussion} shows numerical results and visual interpretation. Finally, Sec. \ref{sec:Conclusion} concludes the paper.

We use lowercase letters for scalars, lowercase bold-faced letters for vectors and uppercase bold-faced letter for matrices. Further, $(\cdot)^\top$ and $(\cdot)^{\rm H}$ are transpose and hermitian transpose operations, respectively. We use $\mathcal{CN}(\cdot,\cdot)$ and $\mathcal{U}(\cdot,\cdot)$ to denote complex Gaussian distribution and uniform distribution. We use $\Re\{\cdot\}$ and $\Im\{\cdot\}$ to denote real and imaginary part of complex scalar. For a matrix $\boldsymbol{A}$, $[\boldsymbol{A}]_{i,j}$ denote $i$-th row and $j$-th column entry of $\boldsymbol{A}$.

\section{System Model and Problem Formulation}
\label{sec:systemModelNProblemFormulation}
\subsection{System Model}
\begin{figure}[t]
  \centering
  \includegraphics[width=0.47\textwidth]{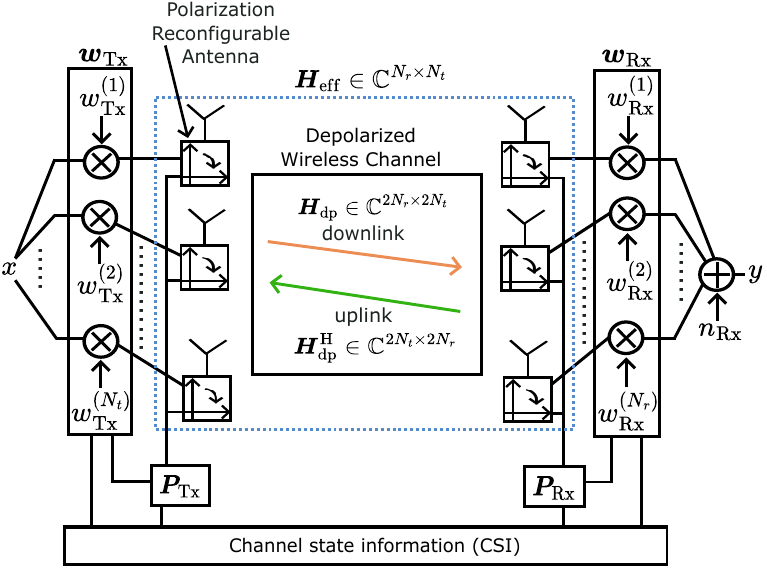} 
  \caption{PR-MIMO system model.}
  \label{fig:PR-MIMO_system_model}
\end{figure}

 This paper considers a point-to-point massive polarization reconfigurable multiple-input-multiple-output (PR-MIMO) communication system, in which a transmitter (Tx) and a receiver (Rx) are equipped with $N_t$ and $N_r$ PR antenna elements, respectively. Both the Tx and Rx are configured as uniform linear arrays (ULA) and each are equipped with a single radio frequency (RF) chain. As depiced in detail in Fig. \ref{fig:PR-MIMO_system_model}, $\boldsymbol{H}_{\rm dp}\in\mathbb{C}^{2N_r\times2N_t}$ denotes the \emph{depolarized} wireless channel matrix from the Tx to the Rx. The polarization matrices, $\boldsymbol{P}_{\rm Tx}$ and $\boldsymbol{P}_{\rm Rx}$ effectively reconfigure $\boldsymbol{H}_{\rm dp}$ as $\boldsymbol{H}_{\rm eff}=\boldsymbol{P}_{\rm Rx}^\top\boldsymbol{H}_{\rm dp}\boldsymbol{P}_{\rm Tx}$ to promote a beneficial wireless environment. The Tx and the Rx polarization matrices, respectively, are defined as
\begin{subequations} \label{eq:BlockdiagPols}
\begin{align}
 \label{eq:TxBlockDiagPols}
    \boldsymbol{P}_{\rm Tx}={\rm blkdiag}(\boldsymbol{p}_{{\rm Tx},1}, \dots, \boldsymbol{p}_{{\rm Tx},N_t}) \in \mathbb{R}^{2N_t\times N_t}, \\
 \label{eq:RxBlockDiagPols}
    \boldsymbol{P}_{\rm Rx}={\rm blkdiag}(\boldsymbol{p}_{{\rm Rx},1}, \dots, \boldsymbol{p}_{{\rm Tx},N_r}) \in \mathbb{R}^{2N_r\times N_r},
\end{align}
\end{subequations} 
where polarization vectors $\boldsymbol{p}_{{\rm Tx},j}$ and $\boldsymbol{p}_{{\rm Rx},i}$ are placed in the diagonal entries of the Tx and Rx polarization matrices, respectively. The polarization vectors at $j$-th Tx and $i$-th Rx PR antenna elements are, respectively,
\begin{subequations} \label{eq:PolVectors}
    \begin{align}
 \label{eq:Tx Polarization vectors}
 \boldsymbol{p}_{{\rm Tx},j} = \left[
    \begin{matrix}
        \cos{\theta_j} & \sin{\theta_j}
    \end{matrix}
    \right]^\top, ~ j=1,\dots,N_t, \\
 \label{eq:Rx Polarization vectors}
    \boldsymbol{p}_{{\rm Rx},i} = \left[
    \begin{matrix}
        \cos{\theta_i} & \sin{\theta_i}
    \end{matrix}
    \right]^\top, ~ i=1, \dots,N_r,
\end{align}
\end{subequations}
where $\theta_j, \theta_i \in [0, \pi/2]$.

A data symbol, $x\in\mathbb{C}$, is defined and assumed to have power constraint $\mathbb{E}[|x|^2]\leq p$. The Tx transmits the data symbol $x$ to the Rx through a normalized transmit beamforming vector $\boldsymbol{w}_{\rm Tx}\in\mathbb{C}^{N_t}$, i.e., $||\boldsymbol{w}_{\rm Tx}||^2=1$; and the Rx receives the symbol through a normalized receive beamforming vector $\boldsymbol{w}_{\rm Rx}\in\mathbb{C}^{N_r}$, i.e, $||\boldsymbol{w}_{\rm Rx}||^2=1$. 
Then, the received baseband signal at the Rx side can be expressed as: 
\begin{equation}
    y = \boldsymbol{w}_{\rm Rx}^{\rm H}\underbrace{\boldsymbol{P}_{\rm Rx}^\top \boldsymbol{H}_{\rm dp} \boldsymbol{P}_{\rm Tx}} _{\boldsymbol{H}_{\rm eff}}\boldsymbol{w}_{\rm Tx}x+n_{\rm Rx}\in \mathbb{C},
\end{equation}
where $n_{\rm Rx}\sim\mathcal{CN}(0,\sigma_{\rm Rx}^2)$ is the additive white Gaussian noise (AWGN) at the Rx. To this end, we express the achievable rate $R$ at the Rx as 
\begin{equation}
 \label{eq:rate}
    R = \log\left(1+\frac{\left|\boldsymbol{w}_{\rm Rx}^{\rm H}\boldsymbol{P}_{\rm Rx}^\top \boldsymbol{H}_{\rm dp} \boldsymbol{P}_{\rm Tx}\boldsymbol{w}_{\rm Tx}\right|^2}{\sigma^2_{\rm Rx}}\right). 
\end{equation}
The objective of this paper is to jointly design polarization matrices $\boldsymbol{P}_{\rm Rx}$ and $\boldsymbol{P}_{\rm Tx}$; beamforming vectors $\boldsymbol{w}_{\rm Rx}$ and $\boldsymbol{w}_{\rm Tx}$, such that the achievable rate $R$ is maximized. We note that the design on $\boldsymbol{P}_{\rm Tx}$ and $\boldsymbol{P}_{\rm Rx}$ is equivalent to designing the polarization angles $\boldsymbol{\theta}_{\rm Tx}=[\theta_1,\dots,\theta_{N_t}]$ and $\boldsymbol{\theta}_{\rm Rx}=[\theta_1,\dots,\theta_{N_r}]$. The problem of interest is then formulated as 
\begin{equation}
 \label{eq:max_formulation}
\begin{aligned}
\max_{\boldsymbol{\theta}_{\rm Tx},\boldsymbol{\theta}_{\rm Rx},\boldsymbol{w}_{\rm Tx}, \boldsymbol{w}_{\rm Rx}} \quad & \left|\boldsymbol{w}_{\rm Rx}^{\rm H}\boldsymbol{P}_{\rm Rx}^\top \boldsymbol{H}_{\rm dp} \boldsymbol{P}_{\rm Tx}\boldsymbol{w}_{\rm Tx}\right|^2\\
\textrm{subject to} \quad & ||\boldsymbol{w}_{\rm Tx}||^2=1, \\
\quad & ||\boldsymbol{w}_{\rm Rx}||^2 = 1, \\
  & (\ref{eq:BlockdiagPols}), (\ref{eq:PolVectors}).\\
\end{aligned}
\end{equation} 

Steps to solve problem (\ref{eq:max_formulation}) with the perfect CSI $\boldsymbol{H}_{\rm dp}$, is described as follows. We define $\boldsymbol{P}_{\rm Tx}^*$ and $\boldsymbol{P}_{\rm Rx}^*$ as the optimal Tx and Rx polarization matrices, respectively. They are obtained by iterative polarization optimization (IPO) algorithm described in \cite{Kwon_Molisch_Globecom, AntennaSelection} which relies on the knowledge of $\boldsymbol{H}_{\rm dp}$. Subsequently, $\boldsymbol{H}_{\rm dp}$ is reconfigured as $\boldsymbol{H}_{\rm eff}^*=(\boldsymbol{P}_{\rm Rx}^*)^\top\boldsymbol{H}_{\rm dp}\boldsymbol{P}_{\rm Tx}^*$, which can be decomposed by singular value decomposition (SVD) as $\boldsymbol{H}_{\rm eff}^*=\boldsymbol{U\Sigma V}^{\rm H}$. The optimal Tx beamforming vector $\boldsymbol{w}^*_{\rm Tx}$ is the normalized singular vector included in the right singular matrix $\boldsymbol{V}$ that corresponds to the largest singular value \cite{cover1999elements}. That is, $\boldsymbol{w}^*_{\rm Tx}=\boldsymbol{v}_{\rm max}/||\boldsymbol{v}_{\rm max}||_2$. In an analogous manner, the optimal Rx beamforming vector $\boldsymbol{w}_{\rm Rx}^*$ is the normalized singular vector of the left singular matrix $\boldsymbol{U}$ that corresponds to the largest singular value, i.e., $\boldsymbol{w}^*_{\rm Tx}=\boldsymbol{u}_{\rm max}/||\boldsymbol{u}_{\rm max}||_2$. 

As described, steps to solve (\ref{eq:max_formulation}) relies on the perfect CSI. However, acquiring the CSI in a large-scale PR-MIMO system with limited RF chain is extremely challenging because both the Tx and Rx can only observe pilots that are lower-dimensional than $\boldsymbol{H}_{\rm dp}$. This is first due to the fact that PR-antenna channel increases the channel dimension by four-fold with depolarized channel matrix $\boldsymbol{H}_{\rm dp}$. Second, the limited RF chains reduce the pilot dimension on both transceiver sides. To tackle this problem, we use the ping-pong pilot protocol with the assumption that the system is operating in the time division duplex (TDD) mode with the uplink-downlink channel reciprocity. 

\subsection{Ping-Pong Pilot Scheme}
This section describes the ping-pong pilot scheme, where the Tx and Rx alternately send pilot symbols while utilizing and updating their parameters, based on sequential information from the accumulated pilots. It is worth emphasizing that during this pilot protocol, each of the Tx and Rx functions as both the transmitter and receiver. In the $l$-th pilot stage, the Tx sends the Rx $l$-th pilot $x_{\textrm{Tx},l}$ under power constraint $\mathbb{E}[|x_{\textrm{Tx},l}|^2]\leq \rho_{\rm Tx}$. Subsequently, the Rx sends the Tx $l$-th pilot $x_{{\rm Rx},l}$ under power constraint $\mathbb{E}[|x_{\textrm{Rx},l}|^2]\leq \rho_{\rm Rx}$. Without loss of generality, we set $x_{\textrm{Tx},l}=\sqrt{\rho_{\rm Tx}}$ and $x_{\textrm{Rx},l}=\sqrt{\rho_{\rm Rx}}$. Channel reciprocity is assumed as portrayed in Fig. \ref{fig:PR-MIMO_system_model}, i.e., we denote the channel from the Tx to the Rx and the channel from the Rx to the Tx as $\boldsymbol{H}_{\rm dp}$ and $\boldsymbol{H}^{\rm H}_{\rm dp}$, respectively. 

Further, the $l$-th received pilot at the Tx is described as
\begin{equation}
 \label{eq:lthTxpilot}
    y_{\textrm{Tx},l}=(\boldsymbol{w}_{\textrm{Tx},l}^{\rm r})^{\rm H}(\boldsymbol{P}_{\textrm{Tx},l}^{\rm r})^\top\boldsymbol{H}_{\rm dp}^{\rm H}\boldsymbol{w}_{\textrm{Rx},l}^{\rm t}\boldsymbol{P}_{\textrm{Rx},l}^{\rm t}x_{\textrm{Tx},l}+n_{{\rm Tx},l},
\end{equation}
where $\boldsymbol{w}_{{\rm Rx},l}^{\rm t}$, $\boldsymbol{P}_{{\rm Rx},l}^{\rm t}$, $\boldsymbol{w}_{{\rm Tx},l}^{\rm r}$ and $\boldsymbol{P}_{{\rm Rx},l}^{\rm r}$ are respectively, the Tx receiving beamforming vector and polarization matrix and the Rx transmitting beamforming vector and polarization matrix used in $l$-th pilot stage. Here, $n_{{\rm Tx},l}\sim\mathcal{CN}(0,\sigma_{\rm Tx}^2)$ is AWGN at the Tx side during $l$-th pilot stage. In an analogous manner, $l$-th received pilot at the receiver side is expressed as
\begin{equation}
 \label{eq:lthRxpilot}
    y_{{\rm Rx},l}=(\boldsymbol{w}^{\rm r}_{{\rm Rx},l})^{\rm H}(\boldsymbol{P}^{\rm r}_{{\rm Rx},l})^\top\boldsymbol{H}_{\rm dp}\boldsymbol{w}^{\rm t}_{{\rm Tx},l}\boldsymbol{P}^{\rm t}_{{\rm Tx},l}x_{\textrm{Rx},l}+n_{{\rm Rx},l}, 
\end{equation}
where $\boldsymbol{w}_{{\rm Rx},l}^{\rm r}$, $\boldsymbol{P}_{{\rm Rx},l}^{\rm r}$, $\boldsymbol{w}_{{\rm Tx},l}^{\rm t}$, $\boldsymbol{P}_{{\rm Tx},l}^{\rm t}$ and $n_{{\rm Rx},l}$ are respectively, the Rx receiving beamforming vector and polarization matrix, the Tx transmitting beamforming vector and polarization matrix and $n_{{\rm Rx},l}\sim\mathcal{CN}(0,\sigma_{{\rm Rx},l}^2)$ is AWGN in $l$-th pilot stage. 
\begin{figure}[t]
  \centering
  \includegraphics[width=0.47\textwidth]{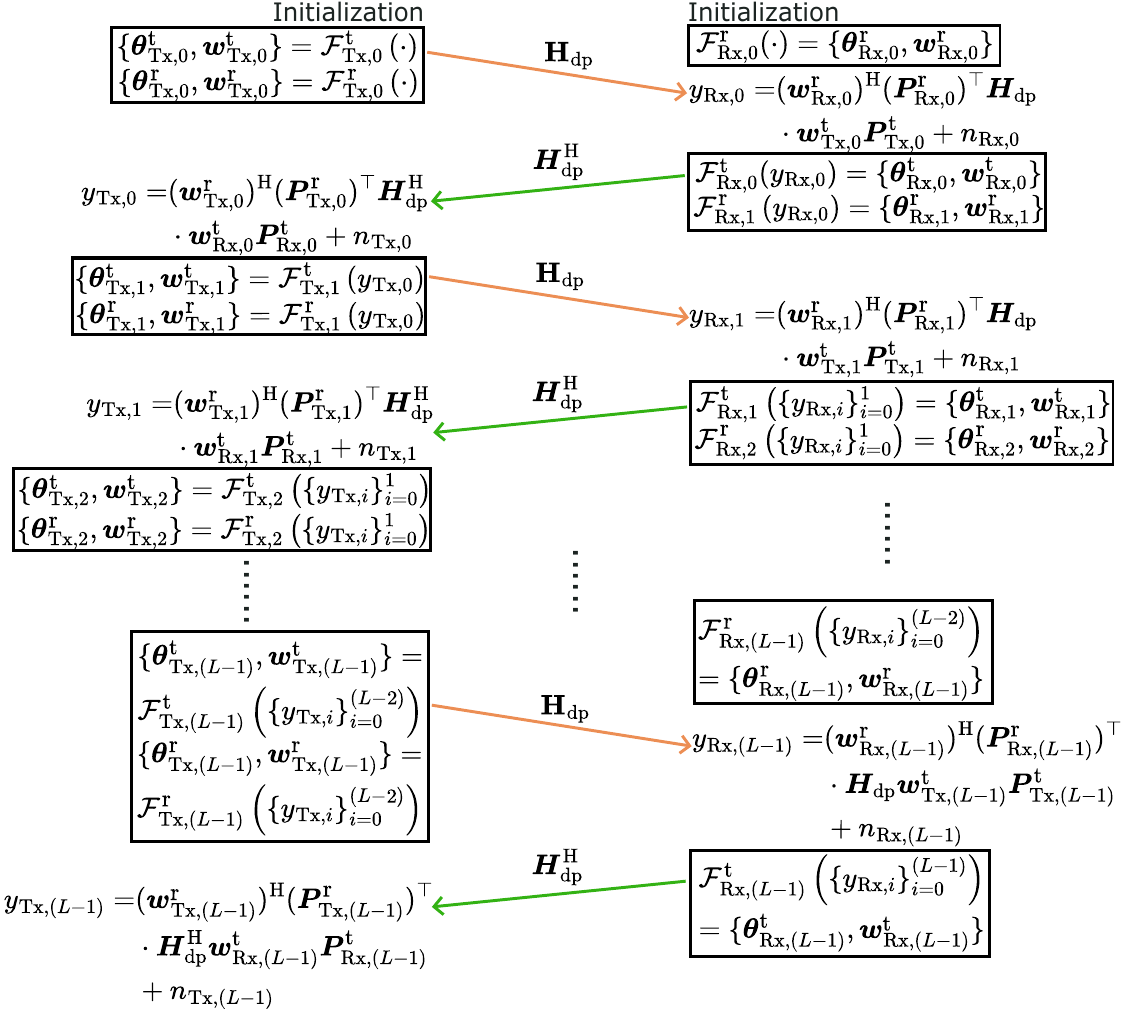} 
  \caption{Ping-pong pilot scheme}
  \label{fig:Pingpong scheme}
\end{figure}

The ping-pong pilot scheme is depicted in detail in Fig. \ref{fig:Pingpong scheme}, where the scheme utilizes temporally correlated information over the accumulated pilot symbols to design parameters required at the next pilot stage for both ends of transceivers. That is, at the $l$-th pilot stage, the Tx transmits a pilot symbol to the Rx using the transmiting polarization matrix $\boldsymbol{P}^{\rm t}_{\textrm{Tx},l}$ updated by polarization angles $\boldsymbol{\theta}^{\rm t}_{\textrm{Tx},l}$ and transmitting beamforming vector $\boldsymbol{w}^{\rm t}_{\textrm{Tx},l}$, which are determined by the accumulated received pilots prior to the $l$-th pilot stage. The process is expressed as
\begin{equation}
 \label{eq:Tx transmitting mapping}
    \{\boldsymbol{\theta}_{{\rm Tx},l}^{\rm t},\boldsymbol{w}_{{\rm Tx},l}^{\rm t}\}=\mathcal{F}_{{\rm Tx},l}^{\rm t}\left(\{y_{{\rm Tx},i}\}_{i=0}^{l-1}\right),
\end{equation}
where $l=0,...,L-1$ and $\mathcal{F}_{{\rm Tx},l}^{\rm t}:\mathbb{C}^{l+1}\rightarrow{\mathbb{C}^{N_t}\times\mathbb{R}^{N_t}}$. 
In a completely analogous manner, the Tx updates the receiving polarization matrix $\boldsymbol{P}^{\rm r}_{\textrm{Tx},l}$ based on polarization angles $\boldsymbol{\theta}^{\rm r}_{\textrm{Tx},l}$ and receiving beamforming vector $\boldsymbol{w}^{\rm r}_{\textrm{Tx},l}$ to receive the $l$-th pilot symbol from the Rx based on $l-1$ pilot symbols received so far. That is,
\begin{equation}
 \label{eq:Tx receiving mapping}
    \{\boldsymbol{\theta}_{{\rm Tx},l}^{\rm r},\boldsymbol{w}_{{\rm Tx},l}^{\rm r}\}=\mathcal{F}_{{\rm Tx},l}^{\rm r}\left(\{y_{\textrm{Tx},i}\}_{i=0}^{l-1}\right),
\end{equation}
where $\mathcal{F}_{{\rm Tx},l}^{\rm r}:\mathbb{C}^{l+1}\rightarrow{\mathbb{C}^{N_t}\times\mathbb{R}^{N_t}}$. 

In turn, the Rx receives the $l$-th pilot from the Tx with receiving polarization matrix $\boldsymbol{P}^{\rm r}_{\textrm{Rx},l}$ from $\boldsymbol{\theta}^{\rm r}_{\textrm{Rx},l}$ and receiving beamforming vector $\boldsymbol{w}^{\rm r}_{\textrm{Rx},l}$ that were updated by $(l-1)$ pilot symbols received pilots, i.e.,
\begin{align}
 \label{eq:Rx receiving mapping}
    \{\boldsymbol{\theta}_{\textrm{Rx},l}^{\rm r},\boldsymbol{w}_{\textrm{Rx},l}^{\rm r}\}=\mathcal{F}_{\textrm{Rx},l}^{\textrm{r}}\left(\{y_{\textrm{Rx},i}\}_{i=0}^{l-1}\right),
\end{align}
where $l=0,\dots,N-2$ and $\mathcal{F}_{\textrm{Rx},l}^{\rm r}:\mathbb{C}^{l+1}\rightarrow{\mathbb{C}^{N_r}\times\mathbb{R}^{N_r}}$. 
In a completely analogous manner again, the Rx transmits the $l$-th pilot symbol to the Tx using transmitting polarization matrix $\boldsymbol{P}_{\textrm{Rx},l}^t$ and transmitting beamforming vector $\boldsymbol{w}_{\textrm{Rx},l}^t$ updated at the end of Rx in Fig. \ref{fig:PR-MIMO_system_model}, based on $l$ pilot symbols received so far. This process is described as
\begin{equation}
 \label{eq:Rx transmitting mapping}
    \{\boldsymbol{\theta}_{\textrm{Rx},l}^{\rm t},\boldsymbol{w}_{\textrm{Rx},l}^{\rm t}\}=\mathcal{F}^{\rm t}_{\textrm{Rx},l}\left(\{y_{\rm Rx,i}\}_{i=0}^{l}\right),
\end{equation}
where $\mathcal{F}^{\rm t}_{\textrm{Rx},l}:\mathbb{C}^{l+1}\rightarrow{\mathbb{C}^{N_r}\times\mathbb{R}^{N_r}}$.

It is noteworthy that all beamforming vectors satisfy the unit norm constraint, i.e., $||\boldsymbol{w}_{\textrm{Tx},l}^{\rm r}||^2=||\boldsymbol{w}_{\textrm{Tx},l}^{\rm t}||^2=||\boldsymbol{w}_{\textrm{Rx},l}^{\rm r}||^2=||\boldsymbol{w}_{\textrm{Rx},l}^{\rm t}||^2=1$ while all polarization angles $\boldsymbol{\theta}_{{\rm Tx},l}^{\rm r}, \boldsymbol{\theta}_{{\rm Tx},l}^{\rm t}, \boldsymbol{\theta}_{{\rm Rx},l}^{\rm r}, \boldsymbol{\theta}_{{\rm Rx},l}^{\rm t}\in [0,\pi/2]$. 

\subsection{Problem Formulation}
At every pilot stage, the Tx and Rx design the downlink parameters for data transmission. In particular, at $l$-th pilot stage, the Tx designs the downlink polarization matrix $\boldsymbol{P}_{\textrm{Tx},l}$ from $\boldsymbol{\theta}_{\textrm{Tx},l}$ and downlink beamforming vector $\boldsymbol{w}_{\textrm{Tx},l}$ for data transmission based on received pilots so far. In similar manner, the Rx designs the downlink polarization matrix $\boldsymbol{P}_{\textrm{Rx},l}$ and downlink beamforming vector $\boldsymbol{w}_{\textrm{Tx},l}$ based on so far observed pilot signals. They are mathematically described as 
\begin{subequations} 
 \label{eq:downlink tranmission}
    \begin{align}
        \{\boldsymbol{\theta}_{\textrm{Tx},l},\boldsymbol{w}_{\textrm{Tx},l}\}=\mathcal{G}_{\textrm{Tx},l}(\{y_{\textrm{Tx},i}\}_{i=0}^{l-1}), \label{eq:downlink for Tx} \\ 
        \{\boldsymbol{\theta}_{\textrm{Rx},l},\boldsymbol{w}_{\textrm{Rx},l}\}=\mathcal{G}_{\textrm{Rx},l}(\{y_{\textrm{Rx},i}\}_{i=0}^{l-1}), \label{eq:downlink for Rx}
    \end{align}
\end{subequations}
where $l=0,\dots,L-1$ and $\mathcal{G}_{\textrm{Tx},l}:\mathbb{C}^{l+1}\rightarrow{\mathbb{C}^{N_t}\times\mathbb{R}^{N_t}}$ and $\mathcal{G}_{\textrm{Rx},l}:\mathbb{C}^{l+1}\rightarrow{\mathbb{C}^{N_r}\times\mathbb{R}^{N_r}}$. The downlink beamforming vectors all satisfy unit norm constraint while downlink polarization angles lie in the range $[0,\pi/2]$. 

The problem of interest is then described as 
\begin{equation}
 \label{eq:max_reformulation}
\begin{aligned}
\max_{\substack{\mathcal{S}}} \quad & \mathbb{E}\left[\frac{1}{L}\sum_{l=0}^{L-1}\left|\boldsymbol{w}_{\textrm{Rx},l}^{\rm H}\boldsymbol{P}_{\textrm{Rx},l}^\top \boldsymbol{H}_{\rm dp} \boldsymbol{P}_{\textrm{Tx},l}\boldsymbol{w}_{\textrm{Tx},l}\right|^2\right]\\
\textrm{subject to} \quad & (\ref{eq:Tx transmitting mapping}), (\ref{eq:Tx receiving mapping}), (\ref{eq:Rx receiving mapping}), (\ref{eq:Rx transmitting mapping}), (\ref{eq:downlink tranmission}), \\
\end{aligned}
\end{equation} 
where the problem is optimized over the set of functions 
\begin{equation}
\begin{aligned}
\mathcal{S}=\Big\{\{\mathcal{F}_{{\rm Tx},i}^{\rm r}(\cdot)\}_{i=0}^{L-2}, 
\{\mathcal{F}_{{\rm Tx},i}^{\rm t}(\cdot)\}_{i=0}^{L-2}, 
\{\mathcal{F}_{{\rm Rx},i}^{\rm r}(\cdot)\}_{i=0}^{L-2}, \\  
\{\mathcal{F}_{{\rm Rx},i}^{\rm t}(\cdot)\}_{i=0}^{L-1},  
\{\mathcal{G}_{\textrm{Tx},i}(\cdot)\}_{i=0}^{L-1}, 
\{\mathcal{G}_{\textrm{Rx},i}(\cdot)\}_{i=1}^{L-1} \Big\}
\end{aligned}
\end{equation}
and expectation is taken over the channel realizations. 
This is a challenging variational optimization problem in which the objective is maximized over the high dimensional function mappings (\ref{eq:Tx transmitting mapping}), (\ref{eq:Tx receiving mapping}), (\ref{eq:Rx receiving mapping}), (\ref{eq:Rx transmitting mapping}) and (\ref{eq:downlink tranmission}). Problem (\ref{eq:max_reformulation}) is sequential because all the parameters are optimized over accumulated received pilot; therefore, we propose to learn the function mapping by transformer-based deep learning architecture. 

\section{Proposed Transformer-Based Solution}
\label{sec:RTF_solution}
\begin{figure*}[t]
  \centering
  \includegraphics[width=0.96 \textwidth]{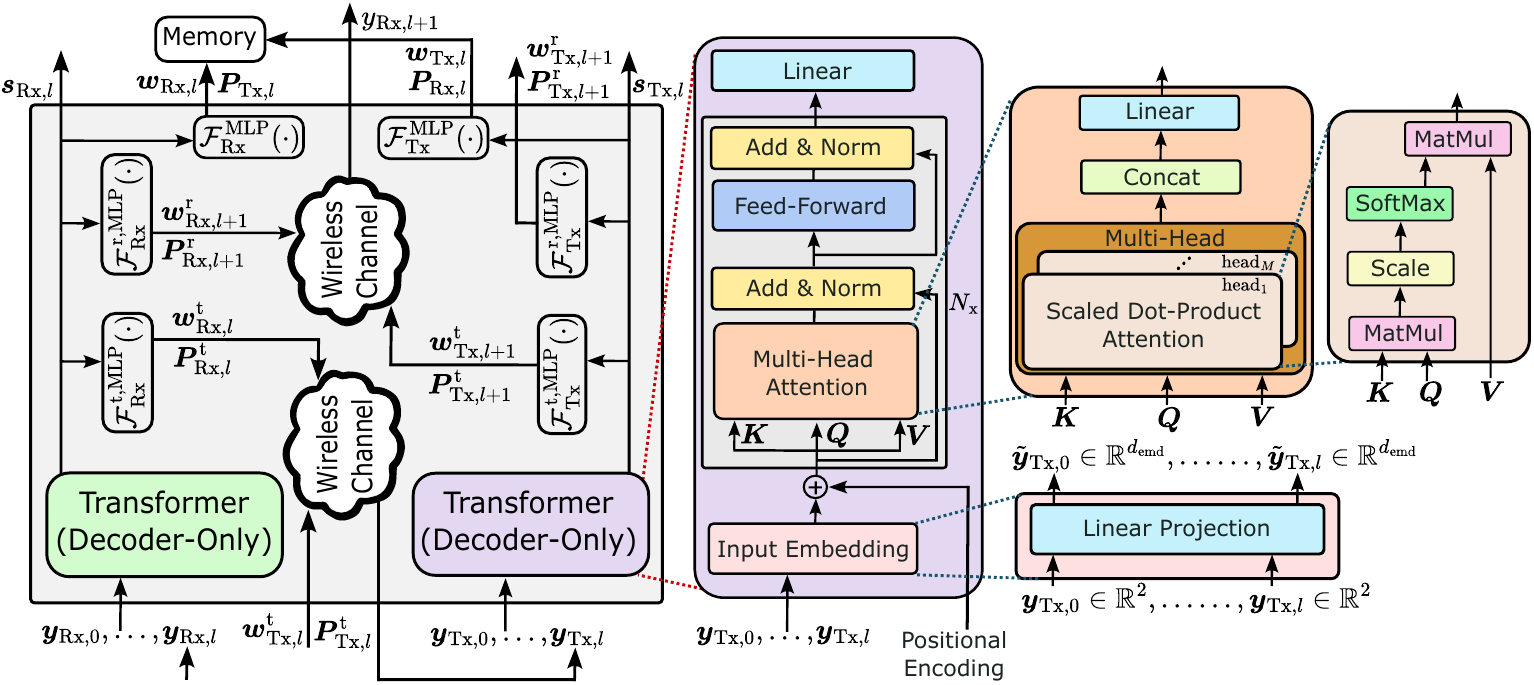} 
  \caption{Proposed transformer unit.}
  \label{fig:Transformer_Arch}
\end{figure*}
This paper proposes a data-driven approach to solve joint PR-BA problem via employing a transformer-based deep learning framework to parameterize the mappings (\ref{eq:Tx transmitting mapping}) -- (\ref{eq:downlink tranmission}) to maximize the beamforming gain in (\ref{eq:max_reformulation}). The fundamental nature of the aforementioned ping-pong pilot scheme is sequential process; consequently, the joint PR-BA problem naturally falls under the realm of problems that can be best solved with transformers. In particular, we leverage the \emph{multi-head attention} mechanism to capture relationship between embedded pilots received at different stages at both ends. By extracting diverse contextual dependencies, multi-head attention enhances the system capability of interpreting pilot signals.


The joint PR-BA problem shares similarities with the generative pretrained transformer (GPT) \cite{GPT}, as it takes a sequence of pilot signals as input and generates pilot parameters for the next sequence. Therefore, we train a decoder-only transformer, similar to a GPT model. Furthermore, the joint PR-BA problem requires both transceiver to optimize their parameters which naturally led us to employ two transformer units on each side to design their respective parameters to maximize the beamforming gain in (\ref{eq:max_formulation}). 

\subsection{Multi-Head Attention Learning Parameters}
\label{subsec:Multi-head Attention}
A transformer captures the temporal correlations of the input sequence via multi-head attention mechanism. This paper proposes to use this ability of the transformer to capture the temporal correlation between the sequence of received pilots on both transceiver ends. As shown by Fig. \ref{fig:Transformer_Arch}, each pilot signals on both ends are projected onto an embedding space, producing $\boldsymbol{K}\in\mathbb{R}^{l\times d_{\rm emb}}$, $\boldsymbol{Q}\in\mathbb{R}^{l\times d_{\rm emb}}$ and value $\boldsymbol{V}\in\mathbb{R}^{l\times d_{\rm emb}}$, which consists of key, query and value, respectively, for each $l$ pilot signals in their rows. Here, $d_{\rm emb}$ is the input embedding dimension. Multi-head attention consists of multiple single-head attention mechanisms operating in parallel. 
Specifically, for each attention head, we use \emph{scaled dot-product} attention \cite{vaswani2017attention} described as 
\begin{equation}
 \label{eq:attention}
    \textrm{Attention}(\boldsymbol{K,Q,V})=\textrm{softmax}\left(\frac{\boldsymbol{QK}^\top}{\sqrt{d_{\rm emb}}}\right)\boldsymbol{V},
\end{equation}
where $\textrm{softmax}\left(\boldsymbol{QK}^\top/\sqrt{d_{\rm emb}}\right)$ produces the \emph{attention score}, which represent the relative importance of each key to the query. These scores are then used to compute a weighted sum of the values $\boldsymbol{V}$, forming the final attention output.
A single attention head is then described as
\begin{equation}
 \label{eq:single_head}
    \textrm{head}_h=\textrm{Attention}(\boldsymbol{K}\boldsymbol{W}^k_h,\boldsymbol{Q}\boldsymbol{W}^q_h,\boldsymbol{V}\boldsymbol{W}^v_h),
\end{equation}
where $\boldsymbol{W}^k_h\in\mathbb{R}^{d_{\rm emb}\times d_h}, \boldsymbol{W}^q_h\in\mathbb{R}^{d_{\rm emb}\times d_h},$ and $\boldsymbol{W}^v_h\in\mathbb{R}^{d_{\rm emb}\times d_h}$ are the trainable parameters for the $h$-th head. Here, $d_h$ is dimension of a single head. Let  $M$ denote the total number of heads in multi-head attention, then we can describe the multi-head attention as follows: 
\begin{equation}
 \label{eq:multi_head}
    \textrm{MultiHead} = \textrm{Concat}(\textrm{head}_1,\dots,\textrm{head}_M)\boldsymbol{W}^o,
\end{equation}
where $\bold{W}^o\in\mathbb{R}^{Md_h\times d_{\rm emb}}$ projects the concatenated attention blocks onto the embedding space. 

As the model trains, it learns to update the following trainable parameters $\left\{ \{\boldsymbol{W}_h^k, \boldsymbol{W}_h^q, \boldsymbol{W}_h^v\}_{h=1}^{M}, \boldsymbol{W}^o\right\},$
for each multi-head attention block in the transformer. These parameters are unique to each transformer layer, meaning that different layers do not share weights. Furthermore, the two separate transformer units at the Tx and Rx employ independent transformers with distinct trainable parameters.

\subsection{Overall Model}
\subsubsection{Initialization} 
As illustrated by Fig. \ref{fig:Transformer_Arch}, the overall transformer-based deep learning framework for joint PR-BA problem is described as follows.
We initialize $\boldsymbol{w}^{\rm t}_{\textrm{Tx},0}, \boldsymbol{w}^{\rm r}_{\textrm{Tx},0}, \boldsymbol{w}^{\rm r}_{\textrm{Rx},0}$ randomly and satisfy the unit norm constraint while we set $\boldsymbol{P}^{\rm t}_{\textrm{Tx},0}, \boldsymbol{P}^{\rm r}_{\textrm{Tx},0}, \boldsymbol{P}^{\rm r}_{\textrm{Rx},0}$ randomly as each polarization angles lies within the region $[0, \pi/2]$.

\subsubsection{Input Embedding and Positional Encoding}
In the $l$-th pilot stage, the Rx takes the vectorized $l$-th pilot signal, i.e., $\boldsymbol{y}_{\textrm{Rx},l}=\left[\Re\{{y}_{\textrm{Rx},l}\},\Im\{{y}_{\textrm{Rx},l}\}\right]\in\mathbb{R}^{2}$, and projects it onto the embedding space. This is done through a linear layer with trainable weights $\boldsymbol{W}_{\rm emb}\in\mathbb{R}^{d_{\rm emb}/2}$ and biases $\boldsymbol{b}_{\text{emb}}\in \mathbb{R}^{d_{\rm emb}/2}$. Each entry of $\boldsymbol{y}_{\textrm{Rx},l}$ is linearly transformed with the $\boldsymbol{W}_{\rm emb}$ and $\boldsymbol{b}_{\rm emb}$ and concatenated, finally producing $\tilde{\boldsymbol{y}}_{\textrm{Rx},l}$. Mathematically this is described as 
\begin{align}
    \tilde{\boldsymbol{y}}_{\textrm{Rx},l} =&
    \textrm{Concat} \Big( 
    \Re\{ y_{\textrm{Rx},l} \} \boldsymbol{W}_{\textrm{emb}} + \boldsymbol{b}_{\textrm{emb}}, \\ \notag
    &\Im\{ y_{\textrm{Rx},l} \} \boldsymbol{W}_{\textrm{emb}} + \boldsymbol{b}_{\textrm{emb}} 
    \Big) \in \mathbb{R}^{d_{\rm emb}}.
\end{align}
After the projection, we concatenate $\tilde{\boldsymbol{y}}_{\textrm{Rx},l}$ with embedded pilot signals prior to $l$-th pilot stage. we describe this as 
$\textrm{Concat}\left(\tilde{\boldsymbol{y}}_{\textrm{Rx},0},\dots,\tilde{\boldsymbol{y}}_{\textrm{Rx},(l-1)}, \tilde{\boldsymbol{y}}_{\textrm{Rx},l}\right)\in \mathbb{R}^{l\times d_{\rm emb}}.$ 

Since transformers lack inherent sequence awareness, we apply positional encoding to the embedded pilot signals, ensuring that sequential information is retained. We adopt the sinusoidal positional encoding from \cite{vaswani2017attention}, defined as:
\begin{subequations} 
 \label{eq:positionalencoding}
    \begin{align}
        \textrm{PE}_{\textrm{pos},2i} &= \sin\left(\frac{\textrm{pos}}{10000^{(2i/d_{\rm emb})}}\right), \label{eq:positionalencoding_a} \\ 
        \textrm{PE}_{\textrm{pos},2i+1} &= \cos\left(\frac{\textrm{pos}}{10000^{(2i/d_{\rm emb})}}\right). \label{eq:positionalencoding_b}
    \end{align}
\end{subequations}

Similarly, at the Tx, the $l$-th pilot signal received from the Rx undergoes the same embedding and positional encoding process. However, the Tx maintains independent trainable parameters for embedding, while still using the shared positional encoding from (\ref{eq:positionalencoding}).

\subsubsection{Transformer Update}
The embedded pilot signals with positional encoding serves as input to the multi-head attention, which operates as described in Sec. \ref{subsec:Multi-head Attention}. After a skip connection and layer normalization, the multi-head attention output is processed by a position-wise feed-forward network (FFN), defined as 
\begin{equation}
    \textrm{FFN}(\boldsymbol{x})=\max(0,\boldsymbol{x}\boldsymbol{W}_{\textrm{ffn},1}+\boldsymbol{b}_{\textrm{ffn},1})\boldsymbol{W}_{\textrm{ffn},2}+\boldsymbol{b}_{\textrm{ffn},2},
\end{equation}
where $\{\boldsymbol{W}_{\textrm{ffn},1},\boldsymbol{W}_{\textrm{ffn},2}\}$ and $\{\boldsymbol{b}_{\textrm{ffn},1},\boldsymbol{b}_{\textrm{ffn},2}\}$ are trainable weights and biases of the FFN. The output of FFN undergoes another skip connection and layer normalization for model stability. 

As shown in Fig. \ref{fig:Transformer_Arch}, the model consists of $N_{\rm x}$ layers of multi-head followed by FFN blocks, each with unique trainable parameters.  After passing through all $N_{\rm x}$ layers, the model outputs the final state representation, i.e., $\boldsymbol{s}_{\textrm{Rx},l}$. Similarly, at each pilot stage, the Tx follows the same process with its own trainable parameters.

\subsubsection{Multi-Layer Perceptrons}
At the Rx side, the updated state $\boldsymbol{s}_{\textrm{Rx},l}$ from the transformer serves as the input to two multi-layer perceptrons (MLPs) described as
\begin{align}
    \boldsymbol{o}_{\rm Rx}^{\rm t}=\mathcal{F}^{\textrm{t,MLP}}_{\rm Rx}(\boldsymbol{s}_{\textrm{Rx},l}) \in \mathbb{R}^{3N_r}, \\ 
    \boldsymbol{o}_{\rm Rx}^{\rm r}=\mathcal{F}^{\textrm{r,MLP}}_{\rm Rx}(\boldsymbol{s}_{\textrm{Rx},l}) \in \mathbb{R}^{3N_r},
\end{align}
where $\mathcal{F}^{\textrm{t,MLP}}_{\textrm{Rx}}(\cdot)$ and $\mathcal{F}^{\textrm{r,MLP}}_{\textrm{Rx}}(\cdot)$ map the state representation $\boldsymbol{s}_{\textrm{Rx},l}$ to the MLPs' outputs. Similarly, at the Tx side, the updated state representation from the transformer is processed through two MLPs, as illustrated by Fig. \ref{fig:Transformer_Arch}. 

\subsubsection{Sigmoid and Normalization Layer}
\label{subsubsection:Sigmoid and Norm}
We process the MLP outputs at the Rx through a sigmoid layer to generate $\boldsymbol{\theta}_{\textrm{Rx},l}^{\rm t}, \boldsymbol{\theta}_{\textrm{Rx},l}^{\rm r}\in [0,\pi/2]$ and normalization layer to generate $\boldsymbol{w}_{\textrm{Rx},l}^t, \boldsymbol{w}_{\textrm{Rx},l}^r$ which satisfy the unit norm constraint. Specifically, to find the transmitting parameters, we let $[\boldsymbol{o}^{\rm t}_{\rm Rx}]_{{d_1}:{d_2}}$ denote a subvector of $\boldsymbol{o}^{\rm t}_{\rm Rx}$ indexed from $d_1$ to $d_2$. The first $[[\boldsymbol{o}^{\rm t}_{\rm Rx}]_{1:N_r}$ are processed through a sigmoid layer:
\begin{equation}
    \boldsymbol{\theta}_{\textrm{Rx},l}^{\rm t}=\sigma([\boldsymbol{o}_{\rm Rx}^{\rm t}]_{1:N_r})\cdot\dfrac{\pi}{2},
\end{equation} 
yielding $N_r$ polarization angles that lie in the range $[0, \pi/2]$, to generate $l$-th pilot stage transmitting pilot polarization matrix at which is constructed according to (\ref{eq:Rx Polarization vectors}). 

For the remaining $2N_r$ elements, we normalize them to obtain the beamforming vector:  
\begin{equation}
    \boldsymbol{w}_{\textrm{Rx},l}^{\rm t}=\frac{[\boldsymbol{o}_{\rm Rx}^{\rm t}]_{{(N_r+1)}:{2N_r}}+[\boldsymbol{o}_{\rm Rx}^{\rm t}]_{{(2N_r+1)}:{3N_r}}}{||[\boldsymbol{o}_{\rm Rx}^{\rm t}]_{{(N_r+1)}:{3N_r}}||_2}.
\end{equation}
This ensures that $\boldsymbol{w}_{\textrm{Rx},l}^{\rm t}$ satisfies the unit norm constraint. Same approach is taken to also find $\boldsymbol{\theta}_{\textrm {Rx},l}^{\rm r}$ and $\boldsymbol{w}_{\textrm{Rx},l}^{\rm r}$.

In an analogous manner, the Tx follows the same procedure with its own independent MLPs to generate polarization matrices and beamforming vectors at each pilot stage, as illustrated in Fig. \ref{fig:Transformer_Arch}.

After every pilot round, we use two more MLPs to output polarization angles and beamforming vectors used for downlink data transmission described 
\begin{align}
    \boldsymbol{o}_{\textrm{Tx},l} = \mathcal{F}^{\rm MLP}_{\textrm{Tx}}(\boldsymbol{s}_{\textrm {Rx},l}) \in \mathbb{R}^{3N_t},\\
    \boldsymbol{o}_{\textrm{Rx},l} = \mathcal{F}_{\rm Rx}^{\textrm{MLP}}(\boldsymbol{s}_{\textrm {Rx},l}) \in \mathbb{R}^{3N_r}. 
\end{align}
They further go through sigmoid and normalization layer to produce downlink polarization and beamforming vector for data transmission. They are stored in the memory to be utilized for loss function.

\subsubsection{Neural Network Training and Inference}
The transformer unit described in Fig. \ref{fig:Transformer_Arch} undergoes the pilot stages and utilize the trained parameters, i.e., weights and biases, inside the model. The proposed tranformer based architecture is trained offline in unsupervised manner with Adam optimizer \cite{adam} which minimizes the loss function defined as 
\begin{equation}
    \label{eq:loss_func}
    loss = -\mathbb{E}\left[\sum_{l=0}^{L-1}\left|\boldsymbol{w}_{{\rm Rx},l}^{\rm H}\boldsymbol{P}_{{\rm Rx},l}^\top \boldsymbol{H}_{\rm dp} \boldsymbol{P}_{{\rm Tx},l}\boldsymbol{w}_{{\rm Tx},l}\right|^2\right].
\end{equation}
Here, $\left\{\boldsymbol{w}_{{\rm Rx},i},\boldsymbol{P}_{{\rm Rx},i}, \boldsymbol{P}_{{\rm Tx},i}\boldsymbol{w}_{{\rm Tx},i}\right\}_{i=0}^{L-1}$ are downlink parameters gathered in memory throughout the $L$ pilot stages and the expectation is approximated by empirical average over the training set. Once the training is complete, the trained transformer unit can be used for inference. 

\section{Simulation Results}
\label{sec:Numerical Experiments}
\subsection{Simulation Setup}
The evaluation of the proposed transformer-based framework takes into account the massive PR-MIMO system with single RF chains on both transceiver ends. The Tx equipped with $N_t=64$ antenna elements supports the Rx equipped with $N_r=32$ antenna elements.
\subsubsection{Channel Model}
It is inevitable to include the component of channel depolarization in the channel model to fully utilize the PR antenna elements, and the simulation in this paper adopts the double-directional channel model \cite{Double_Direction}. As portrayed in Fig. \ref{fig:PR-MIMO_system_model}, PR wireless channel is characterized by the \emph{depolarized} channel matrix $\boldsymbol{H}_{\rm dp}$ modeled as
\begin{equation}
 \label{eq:depolarizedChannel}
 \boldsymbol{H}_{\rm dp}=\sum_{p=1}^{P}\bar{\boldsymbol{J}}_{\textrm{r},p}^\top\left((\beta_p\boldsymbol{a}_{\textrm{r}}(\phi_p^{\textrm{r}})\boldsymbol{a}_{\textrm{t}}^{\top}(\phi_p^{\textrm{t}}))\otimes\boldsymbol{Q}(\psi)\boldsymbol{X}_p\right)\bar{\boldsymbol{J}}_{\textrm{t},p},
\end{equation}
where $P$ is the number of paths from the Tx to the Rx; $\beta_p\sim\mathcal{CN}(0,1)$ is complex path gain of $p$-th path; $\phi^{\textrm{r}}_p$ is the angle of arrival (AoA) for $p$-th path to the Rx; $\phi^t_p$ is the angle of departure (AoD) for $p$-th path from the Tx; and $\boldsymbol{a}_{\textrm{t}}(\cdot)$ and $\boldsymbol{a}_{\textrm{r}}(\cdot)$ are steering vectors with half-wavelength antenna element spacing at the Tx and Rx, respeectively. Steering vectors are expressed as  
\begin{align}
 \label{eq:Steering Vectors Tx}
    \boldsymbol{a}_{\textrm{t}}(\phi^{\textrm{t}}_p)=\left[
    \begin{matrix}
     1 & e^{-j\pi\sin(\phi^{\textrm{t}}_p)} & \dots & e^{-j\pi(N_t-1)\sin(\phi^{\textrm{t}}_p)}
    \end{matrix}
    \right], \\ 
 \label{eq:Steering Vector Rx}
    \boldsymbol{a}_{\textrm{r}}(\phi^{\textrm{r}}_p)=\left[
    \begin{matrix}
     1 & e^{-j\pi\sin(\phi^{\textrm{r}}_p)} & \dots & e^{-j\pi(N_r-1)\sin(\phi^{\textrm{r}}_p)}
    \end{matrix}
    \right].
\end{align}
Further, $\psi$ is rotation angle between the local polarization coordinates at the Tx and the local polarization coordinates at the Rx, and the rotation matrix $\boldsymbol{Q}(\psi)$ is 
\begin{equation}
 \label{eq:rotational matrix}
    \boldsymbol{Q}(\psi)=\left[
    \begin{matrix}
        \cos{(\psi)} & -\sin{(\psi)} \\ 
        \sin{(\psi)} & \cos{(\psi)}
    \end{matrix}
    \right].   
\end{equation} The rotation matrix transforms the transmitting polarization state to be aligned with the rotated antenna orientation at the Rx. Further, $\bar{\boldsymbol{J}}_{\textrm{t},p}$ and $\bar{\boldsymbol{J}}_{\textrm{r},p}$, respectively, denote block diagonal matrix containing antenna gain matrices along the entries on its diagonal at the Tx and Rx, i.e., 
\begin{align}
 \label{eq:block_diag_antenna_gain}
    \bar{\boldsymbol{J}}_{\textrm{t},p}={\rm blkdiag}\left(\boldsymbol{J}_{\textrm{t},1},\dots,\boldsymbol{J}_{\textrm{t},N_t}\right), \\ 
    \bar{\boldsymbol{J}}_{\textrm{r},p}={\rm blkdiag}(\boldsymbol{J}_{\textrm{r},1},\dots,\boldsymbol{J}_{\textrm{r},N_r}).
\end{align}
The antenna gain matrix $\boldsymbol{J}_{\textrm{t},j}$ for the $j$-th antenna of the Tx is
\begin{equation}
 \label{eq:antenna_gain Matrix}
    \boldsymbol{J}_{\textrm{t},j}=\left[
    \begin{matrix}
        \mathcal{G}_{C,p} & \mathcal{G}_{X,p} \\ 
        \mathcal{G}_{X,p} & -\mathcal{G}_{X,p}
    \end{matrix}
    \right],
\end{equation}
where $\mathcal{G}_{C,p}$ and $\mathcal{G}_{X,p}$ are co-polarization and cross-polarization gain, respectively, in the $p$-th path. 
Lastly, $\boldsymbol{X}_p\in \mathbb{C}^{2\times2}$ is the depolarization matrix in the $p$-th path as 
\begin{equation}
\label{eq:depolarization matrix}
    \boldsymbol{X}_p=\sqrt{\frac{1}{1+\chi}}\left[
    \begin{matrix}
        e^{j\alpha_p^{\rm HH}} & \sqrt{\chi}e^{j\alpha_p^{\rm HV}} \\ 
        \sqrt{\chi}e^{j\alpha_p^{\rm VH}} & e^{j\alpha_p^{\rm VV}}
    \end{matrix}
    \right],
\end{equation}
where $\alpha_{p}^{\rm AB}\sim\mathcal{U}(0,\pi)$ is the phase change between polarization A at the Tx and polarization B at the Rx, and $\chi$ is the inverse cross-polarization discrimination (XPD) \cite{Kwon_Stuber_GeoTheory}.
\subsubsection{Channel Simulation Setting Details}
In the realizations of polarized channels, we define the inverse XPD of antennas $\chi_{\rm ant}$ that is set as $\chi_{\rm ant}=0.3$; inverse XPD of the depolarization matrix as $\chi=0.2$.
The co-polariztion and cross-polariation gain is computed as $\mathcal{G}_{C,p}=\frac{1}{1+\chi_{\rm ant}}$ and $\mathcal{G}_{X,p}=\frac{\chi_{\rm ant}}{{1+\chi_{\rm ant}}}$.
For comprehensive neural network training and inference, we set the SNR to be 0 dB from which we can analize and compare the SNR gain.

\subsection{Model Implementation Details}
The proposed transformer-based framework is implemented on PyTorch \cite{PyTorch}. The embedding space dimension and total number of heads are set as $d_{\rm emb}=320$ and $M=5$, respectively. This sets each single head dimension as $d_{h}=64$. Based on this, the dimension of each attention head $\{\boldsymbol{W}^k_h,\boldsymbol{W}^q_h, \boldsymbol{W}^v_h\}_{h=1}^M$ in the transformer is $320 \times 64$. Hence, the projection matrix $\boldsymbol{W}^o$ has the dimension, $320\times320$. The size of $\textrm{FFN}$ inside the transformer block is $320\times640\times320$. Finally, the MLPs for the Tx, $\mathcal{F}^{\rm t,MLP}_{\textrm{Tx}}(\cdot)$, $\mathcal{F}^{\rm r,MLP}_{\textrm{Tx}}(\cdot)$, $\mathcal{F}^{\rm MLP}_{\textrm{Tx}}(\cdot)$ have the dimension, $320\times512\times512\times3N_t$; while the dimension of MLPs for the Rx, $\mathcal{F}^{\rm t,MLP}_{\textrm{Rx}}(\cdot)$, $\mathcal{F}^{\rm r,MLP}_{\textrm{Rx}}(\cdot)$ and $\mathcal{F}^{\rm MLP}_{\textrm{Rx}}(\cdot)$ is $320\times512\times512\times3N_r$. 

Each neuron in the hidden layer of MLPs is activated by rectified linear unit (ReLU) followed by layer normalization. In one epoch, the neural network samples 100 mini-batches, each with size $1024$, which are randomly generated channel realizations modeled as (\ref{eq:depolarizedChannel}). Further, the network parameters are updated with Adam optimizer minimizing (\ref{eq:loss_func}), and we use exponential learning rate \cite{li2019exponential} described as $\eta_t=\eta_0\cdot\gamma^t$, where the initial learning rate $\eta_0$ and the decay factor $\gamma$ are set as $\eta_0 = 10^{-4}$ and $\gamma=0.9999$, respectively; $t$ is the current time step. The network is trained in an unsupervised manner, since the training is stopped when further improvement is not achieved over the loss for 25 epochs. 

\subsection{Benchmarks}
Comprehensive simulation results with three benchmarks are included for thorough performance comparison.
\subsubsection{Benchmark 1. Perfect CSI with IPO and Optimal Beamformers}
The perfect CSI $\boldsymbol{H}_{\rm dp}$ is provided, and we use optimal polarization matrices, $\boldsymbol{P}_{\rm Tx}^*$ and $\boldsymbol{P}_{\rm Rx}^*$, designed by IPO scheme proposed in \cite{Kwon_Molisch_Globecom, AntennaSelection} to reconfigure the $\boldsymbol{H}_{\rm dp}$ as $\boldsymbol{H}^*_{\rm eff}=(\boldsymbol{P}_{\rm Rx}^*)^\top\boldsymbol{H}_{\rm dp}\boldsymbol{P}_{\rm Tx}$. Then we use the optimal beamforming vectors, $\boldsymbol{w}^*_{\rm Tx
}$ and $\boldsymbol{w}^*_{\rm Rx}$, which are respectively, the right and left singular vectors corresponding to the maximum singular value of $\boldsymbol{H}_{\rm eff}$.
\subsubsection{Benchmark 2. Gated Recurrent Unit (GRU)}
This benchmark employs active ping-pong scheme with GRU-based RNN to solve the joint PR-BA problem. Specifically, it employs two separate GRU units at the Tx and Rx to continuously update the temporal correlation among the sequence of received pilots in their hidden state vectors. These hidden state vectors are mapped to polarization and beamforming vectors by MLPs in each pilot stage. The size of hidden state vector in the GRUs is set to be $256$ and MLPs at the Tx and Rx share the same size as the transformer framework designed in this paper. 
\subsubsection{Benchmark 3. Non-adaptive design with DNN \cite{DNN_PMISO}}
This baseline employs non-adaptive method to solve the joint PR-BA problem. The initial polarization matrices and beamforming vectors are randomly set, and fixed during the whole pilot stages. After each transceiver 
receives $L$ pilot symbols, it concatenates the sequence of pilot symbols and maps it to the optimal downlink polarization matrices and beamforming vectors via MLPs. The demension of MLP at the Tx side is $2L\times 512\times 512\times3N_t$, and the Rx side has the dimension $2L\times 512\times 512\times3N_r$.

\subsection{Results}
The primary focus of performance evaluation is on the polarization reconfigurable beamforming gain achieved by the proposed transformer-based framework with the ping-pong pilot scheme pursuing joint PR-BA .

\begin{figure}[t]
    \includegraphics[width=0.45 \textwidth]{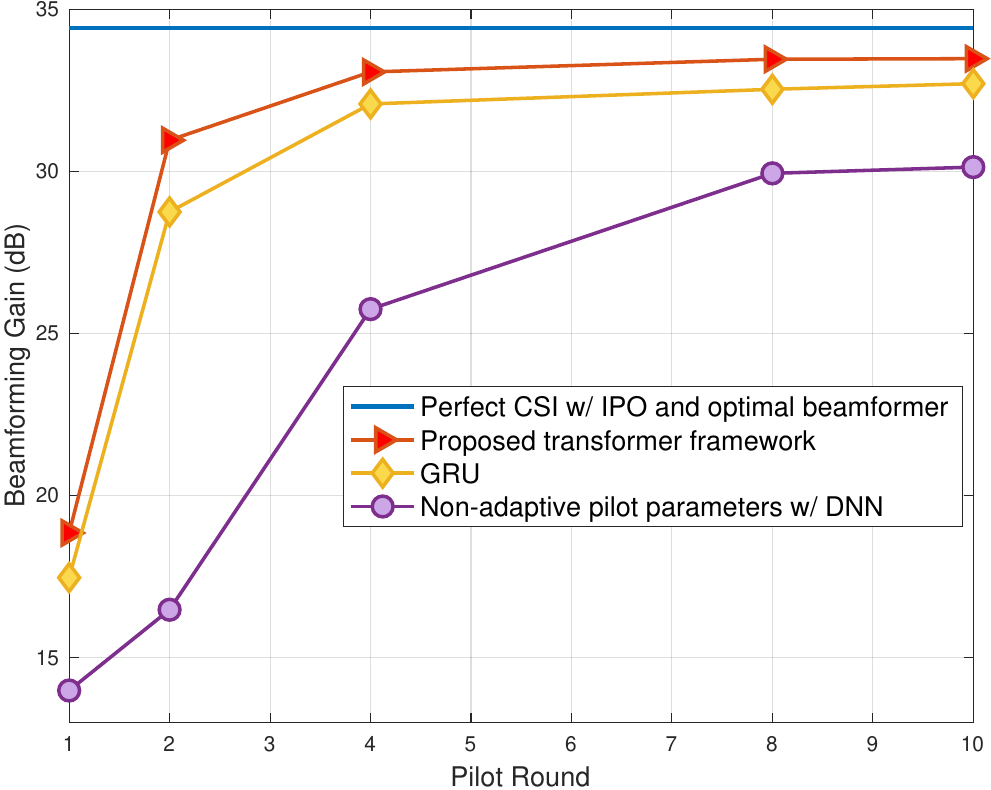}
    \caption{Average beamforming gain vs pilot round ($L$), $P=1$}
    \label{fig:path1_BeamformingGain} 
\end{figure}

\begin{figure}[t]
   \includegraphics[width=0.45 \textwidth]{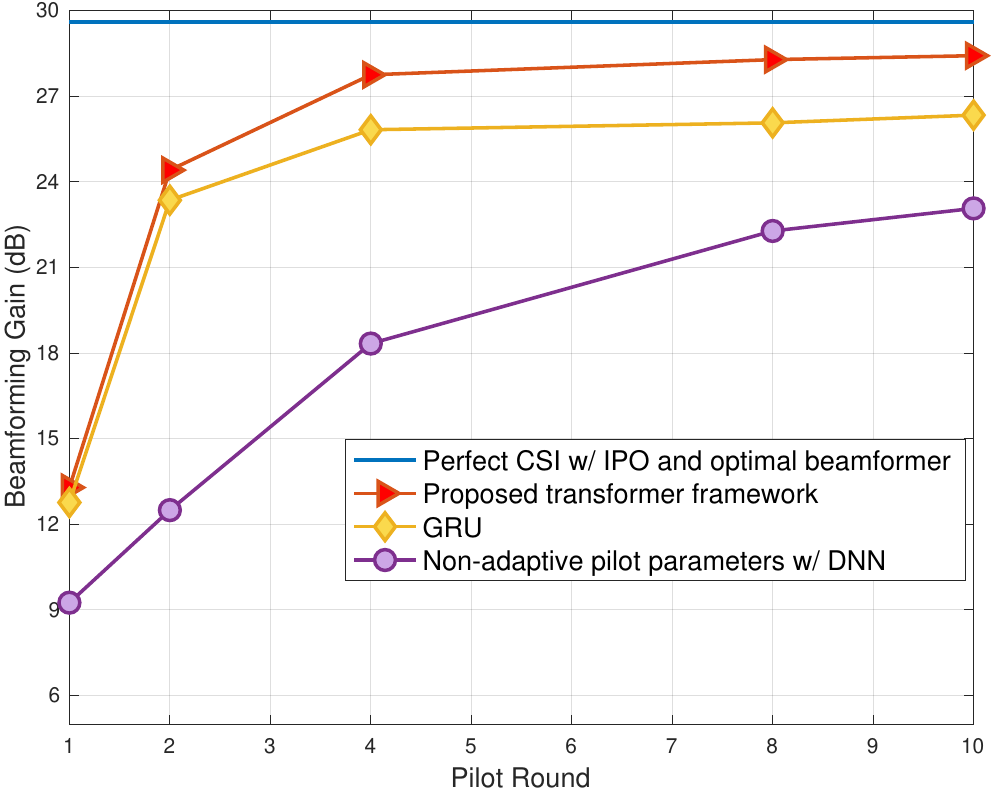}
   \caption{Average beamforming gain vs pilot round ($L$), $P=3$}
   \label{fig:path3_BeamformingGain}
\end{figure}

Figs. \ref{fig:path1_BeamformingGain} and \ref{fig:path3_BeamformingGain} provide the comparison of joint PR-BA beamforming gain with the aforementioned benchmarks in the scenarios of $P=1$ and $P=3$ in (\ref{eq:depolarizedChannel}), respectively.
Simulations have $L$ number of overall pilot stages with $2L$ number of pilot symbols in total. The PR-BA beamforming gain at each pilot stage is averaged over $10^4$ channel realizations. First, both the proposed transformer-based framework and Benchmark 2 with the active pilot scheme have significant gain over Benchmark 3, non-adaptive design.

On the other hand, Figs. \ref{fig:path1_BeamformingGain} and \ref{fig:path3_BeamformingGain} illustrate that the proposed framework consistently achieves a higher average PR-BA beamforming gain than Benchmark 2 (GRU). The average PR-BA beamforming gain of both scheme saturates since $L=4$ in both figures; the transformer-based framework maintains a 1 dB gain and 2 dB gain over Benchmark 2 in Figs. \ref{fig:path1_BeamformingGain} and \ref{fig:path3_BeamformingGain}, respectively. The proposed framework shows higher performance in learning and adapting to the more complex channel environments than the GRU-based framework.

\begin{figure}[t]
   \includegraphics[width=0.45 \textwidth]{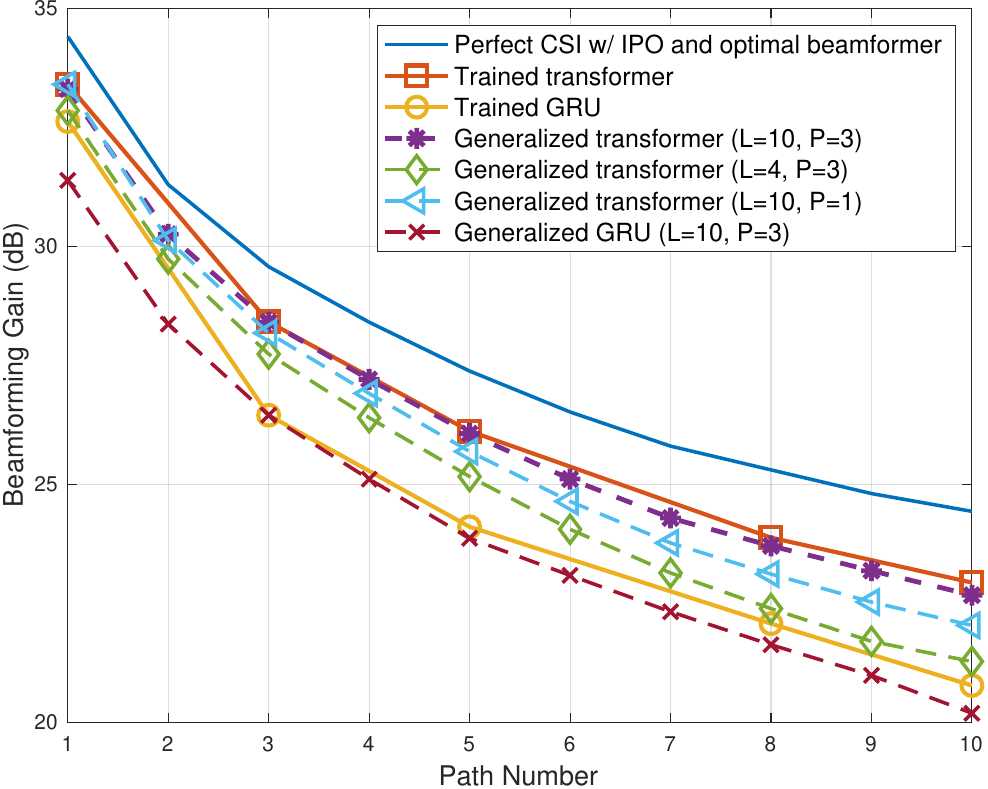}
   \caption{Average beamforming gain vs multi-path, with $L=10$}
   \label{fig:multipath}
\end{figure}

The average PR-BA beamforming gain curves resulted by the proposed transformer-based framework and GRU-based one are compared for a varying number of paths in Fig. \ref{fig:multipath}. The former (orange curve with square marks) outperforms the latter (yellow curve with circle marks) by 2 dB gain at every different number of paths, i.e., $P=1,3,5,8$ and $10$ for the pilot stage, $L=10$.

Several average PR-BA beamforming gain curves (i.e., dashed curves) obtained by generalized transformer are also depicted in Fig. \ref{fig:multipath} for different number of paths and pilot stages. Transformer trained only with $L=10$ and $P=3$; and generalized for all points of the number of paths (purple curve with asterisk marks), exhibits almost the same performance as that of trnasformer trained at each and every points of the number of paths (orange curve with square marks). It is worth mentioning that all generalized transformer models outperform the GRU trained at each and every points, showcasing the robustness of the transformer-based framework. These results highlight that explicit training for every points of the number of paths is unnecessary.

\section{Model Interpretation and Discussion}
\label{sec:InterpretationandDiscussion}
Numerical results presented in Sec. \ref{sec:Numerical Experiments} validate the significant impact of adopting the transformer on the outstanding performance of the proposed scheme. This section interprets the process of the proposed framework and analyzes the reason of the impressive performance. To achieve this, we examine two interpretable outputs from the model: the attention and updated array response.

\subsection{Performance Gain from Multi-Head Attention}
\begin{figure}[t]
  \centering
  \includegraphics[width=0.45 \textwidth]{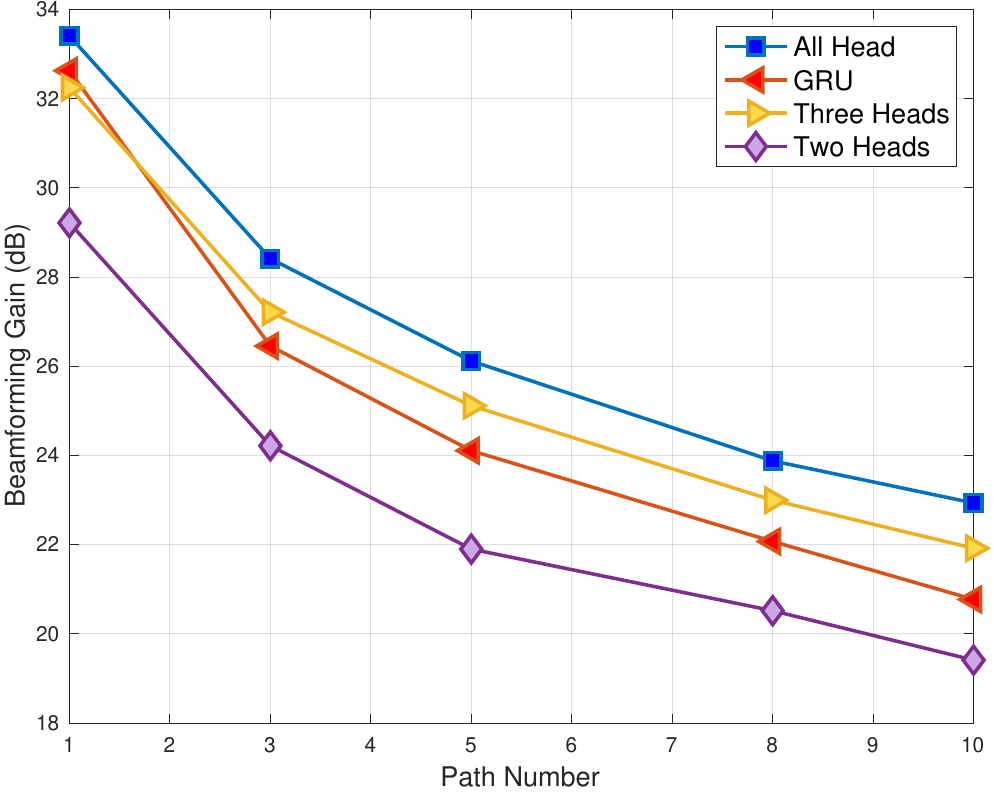} 
  \caption{Average beamforming gain vs multi-path with different number of heads}
  \label{fig:BFHeadL4}
\end{figure}


One key reason for the superior performance of our proposed transformer-based framework over the GRU-based RNN lies in the multi-head attention mechanism. Unlike GRU, which compresses the sequential input pilot signals into a single summary representation, the multi-head attention mechanism allows each attention head to capture diverse contextual information from the input sequence. This enables the transformer to learn a richer representation of the pilot signals, leading to better optimization of polarization and beamforming vectors. In this subsection, we support this claim by analyzing the average beamforming gain achieved with different numbers of attention heads.

To interpret the results from specific heads or subsets of heads within the multi-head attention mechanism, we isolate the heads of interest as follows. From (\ref{eq:multi_head}), we select the desired heads and concatenate them. Then, we modify the output projection matrix $\boldsymbol{W}^o$ by truncating it such that only the rows corresponding to the selected heads are retained. This truncated projection matrix is then applied to the selected attention heads. Mathematically, truncating the projection matrix for $h$-th head is described as
\begin{equation} \label{eq:truncated_projection} \boldsymbol{W}^o_k = \left[\boldsymbol{W}^o\right]_{h d_h:(h+1) d_h,:} \in \mathbb{R}^{{dh} \times d_{\rm model}}, \end{equation}
where $h$ is the head index, and $d_h$ represents the dimensionality of each attention head.

Fig. \ref{fig:BFHeadL4} illustrates the average beamforming gain as a function of the number of multi-path components for different numbers of attention heads. Initially, a subset of heads is randomly selected, and the selected heads are then used to compute the beamforming gain, which is averaged over 10,000 channel realizations. As shown in the figure, Benchmark 2 achieves a higher beamforming gain than the proposed transformer-based framework when only two heads are used. However, as the number of heads increases to three, the proposed transformer-based method surpasses Benchmark 2, achieving approximately 1 dB of additional gain. This result highlights the advantage of multi-head attention, as multiple heads collaboratively extract meaningful information from the received pilot signals.

\subsection{Array Response}
In this section, we use the array response of the Tx to interpret the solution that the model outputs. Since it is crucial to observe the joint impact of polarization and beamforming vectors in our work, we adopt the array response matrix which takes into account for the phase, gain pattern, and polarization differences \cite{Double_Direction}. The Tx array net response matrix is described as:
\begin{equation}
 \label{eq:ANet}
    \boldsymbol{A}^{\rm net}_{\rm Tx}(\theta)=\{\boldsymbol{a}_{\rm t}(\theta)\otimes \left[\begin{matrix} 1 & 1 \end{matrix}\right]\}\odot\boldsymbol{P}_{\rm Tx}^\top\left[
    \begin{matrix}
        \boldsymbol{J}_1 \dots \boldsymbol{J}_{N}
    \end{matrix}\right]^\top\in \mathbb{R}^{N_t\times2},
\end{equation}
where the co-polarized array response is associated in the first column and cross-polarized array response is in the second column. In particular, we use the co-polarized response to observe the array response. Mathematically, the co-polarized part of (\ref{eq:ANet}) is described as 
\begin{equation}
    \label{eq:copolArray}
    \boldsymbol{a}_{\textrm{Tx}}^{\rm C}(\theta) = \left[\boldsymbol{A}_{\textrm{Tx}}^{\rm net}(\theta)\right]_{:,1}. 
\end{equation}
The co-polarized array response of polarization matrix $\boldsymbol{P}_{\rm Tx}$ and beamforming vector $\boldsymbol{w}_{\rm Tx}$ is then described as  
\begin{equation}
 \label{eq:array_response}
    r=\left|(\boldsymbol{w}_{\rm Tx})^{\rm H}\boldsymbol{a}_{\textrm{Tx}}^{\rm C}(\theta)\right|^2, ~ \forall\theta\in[-\pi/2,\pi/2].
\end{equation}

\begin{figure*}
   \centering 
   \includegraphics[width=1\linewidth]{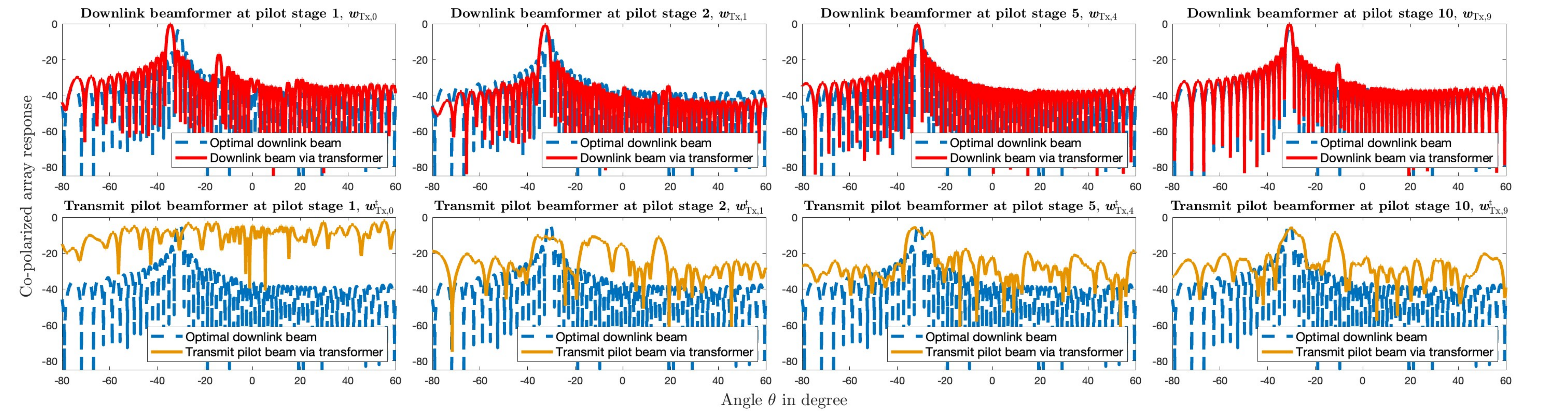}
   \caption{Array response of downlink parameters and transmit pilot parameters via proposed transformer-based framework}
   \label{fig:beampattern of RTF} 
\end{figure*}

\begin{figure*}
   \centering
   \includegraphics[width=1\linewidth]{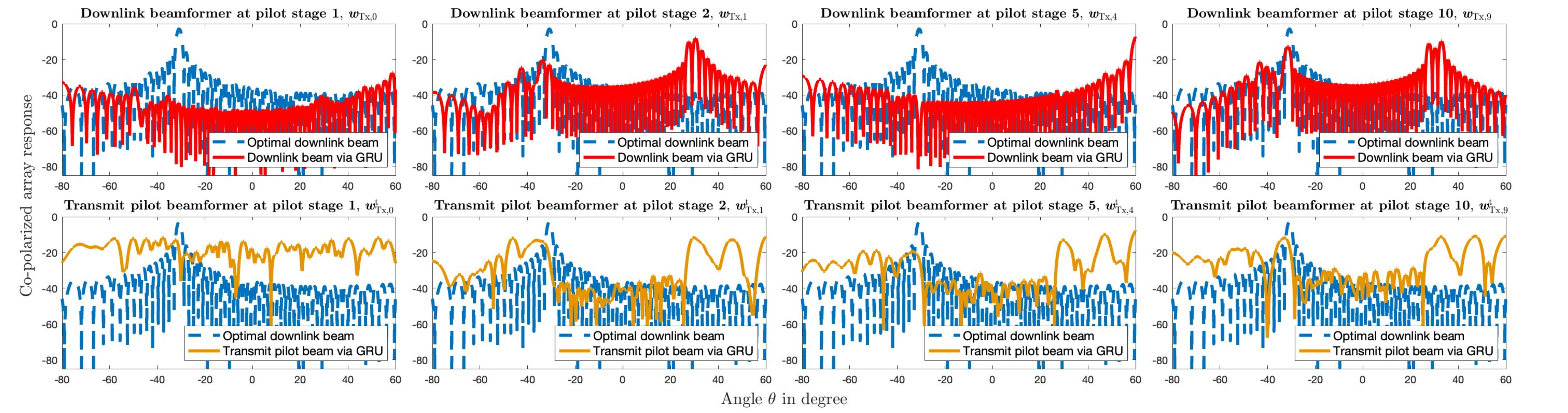}
   \caption{Array response of downlink parameters and transmit pilot parameters via GRU}
   \label{fig:beampattern of GRU}
\end{figure*}



\begin{figure*}
   \includegraphics[width=1\linewidth]{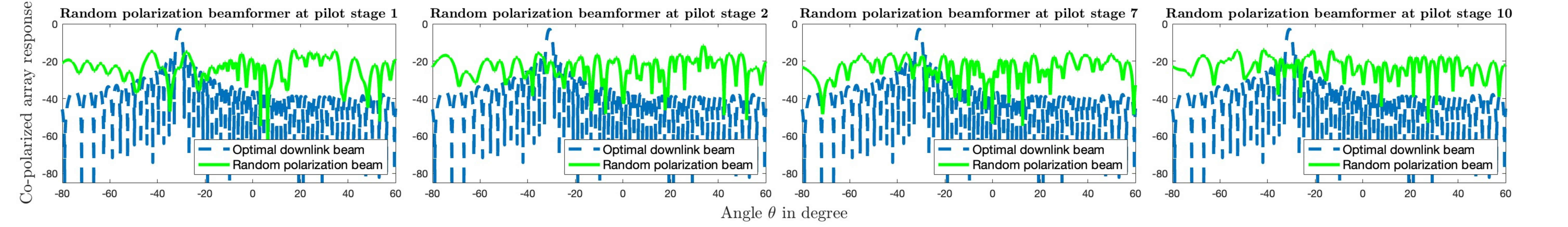}
   \caption{Array response via Random Polarization}
   \label{fig:beampattern of random polarization}
\end{figure*}

In the first row of Fig. \ref{fig:beampattern of RTF}, we plot the array responses of the downlink polarization matrices and beamforming vectors $\boldsymbol{P}_{\textrm {Tx},l}$ and $\boldsymbol{w}_{\textrm{Tx},l}$ designed via the proposed transformer framework for when $l=0,1,6,9$ with a single randomly generated deterministic channel. The neural network used for inference was trained for $L=10$ pilot rounds. We compare this array response with the optimal array response generated by Benchmark 1 (perfect CSI). As illustrated in first row of Fig. \ref{fig:beampattern of RTF}, the downlink array response generated from the proposed transformer-based framework does not align with the optimal array response in the $1$-st pilot stage. However, as the pilot stage proceeds, the transformer is able to gradually refine the beam towards the optimal array response. At last pilot stage, the array response from the transformer finds the main lobe of the optimal array response found via Benchmark 1. 

The behavior of the downlink array response can be further interpreted via the second row of Fig. \ref{fig:beampattern of RTF}. This illustrates the array response of the transmit polarization $\boldsymbol{P}_{\textrm {Tx},l}^{\rm t}$ and beamforming vectors $\boldsymbol{w}_{\textrm {Tx},l}^{\rm t}$ during the pilot stages $l=0,1,6,9$, designed with the proposed transformer framework under the same deterministic channel as first row of Fig. \ref{fig:beampattern of RTF}. As shown, the array response at the first pilot round observes the all direction uniformly. Then, as the round of the pilot proceeds, it gradually refines its direction towards the main beam direction. We observe that it still has a dominant sensing direction around $-15^{o}$ at when pilot stage is 10. This is because the functionality of the polarization and beamforming vectors during pilot stage is to search for directions rather than focusing in one dominant direction. 

We compare the array responses of Fig. \ref{fig:beampattern of RTF} to Fig \ref{fig:beampattern of GRU}, where in the first row, it shows the array response of the downlink polarization and beamforming vectors designed via Benchmark 2 (GRU) under the same channel as Fig. \ref{fig:beampattern of RTF}. As shown, even though the downlink array response is gradually aligning towards the main beam direction of optimal array response, it fails to find the main direction even at the end of $10$-th pilot stage. Similarly, array response of the polarization matrix and beamforming vector during pilot stage, also fails to gradually update its array response towards the main direction. Furthermore, it shows that it is has higher lobes towards $35^{o}$, which ultimately affects the downlink array response to point toward $35^{o}$.  

\begin{figure*}[t]
\centering
\begin{subfigure}[b]{0.32\textwidth} 
    \centering
    \includegraphics[width=\linewidth]{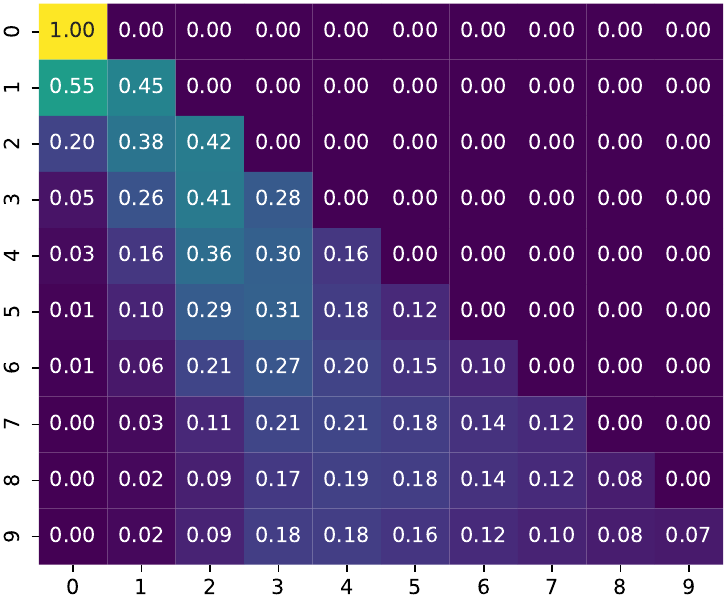}
    \caption{Attention score of head 1}
    \label{fig:heatmap1} 
\end{subfigure}
\hfill
\begin{subfigure}[b]{0.32\textwidth} 
    \centering
    \includegraphics[width=\linewidth]{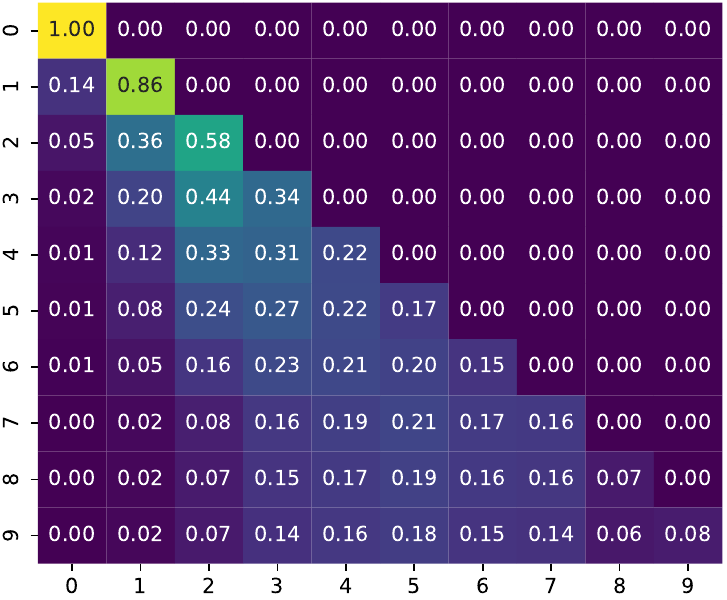}
    \caption{Attention score of head 3}
    \label{fig:heatmap3} 
\end{subfigure}
\hfill
\begin{subfigure}[b]{0.32\textwidth} 
    \centering
    \includegraphics[width=\linewidth]{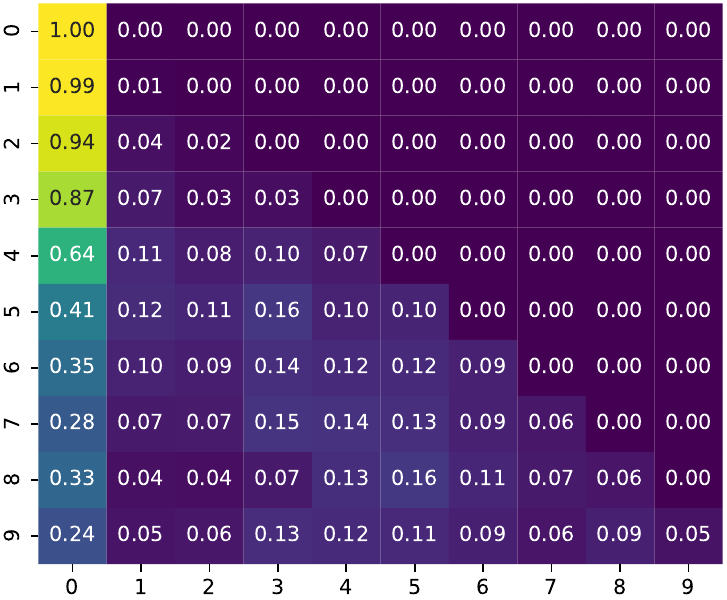}
    \caption{Attention score of head 4}
    \label{fig:heatmap4} 
\end{subfigure}
\vspace{0.3cm} 
\begin{subfigure}[b]{1.0\textwidth} 
    \centering
    \includegraphics[width=\linewidth]{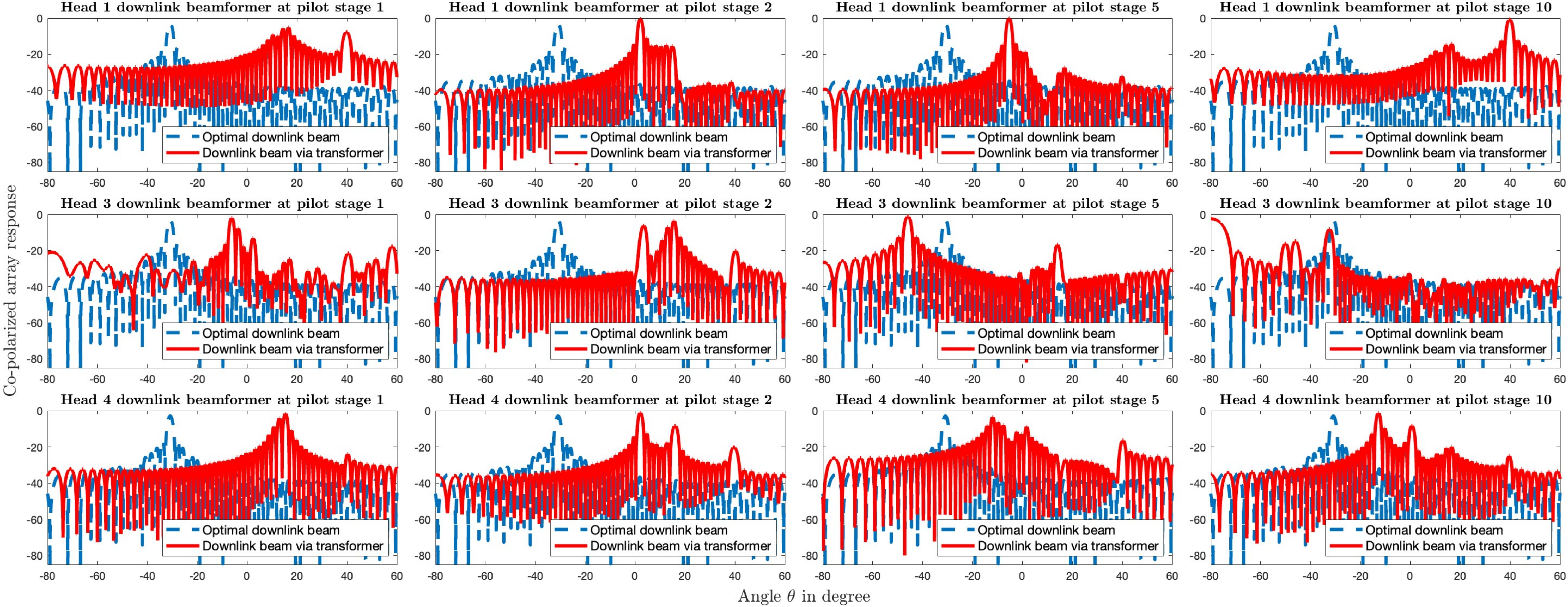}
    \caption{Array response via head 1, head 3, and head 4 at pilot stages 1, 2, 5 and 10}
    \label{fig:beampattern of different attention head}
\end{subfigure}

\caption{Array Response for a specific channel realization $N_t=64, N_r=32, L=10, P=3$}
\label{fig:main}
\end{figure*}

In Fig. \ref{fig:beampattern of random polarization}, we observe the impact of joint optimization design of the proposed transformer framework. Under the same deterministic channel as Fig. \ref{fig:beampattern of RTF}, the array response of randomly generated polarization and optimal downlink beamforming vector is shown in Fig. \ref{fig:beampattern of random polarization}. As illustrated, even though the optimal beamforming vector is well designed, it fails to converge to anything meaningful if it does not work together on top of the optimal polarization vector. 

\subsection{Attention Score}
As explained in Sec. \ref{subsec:Multi-head Attention}, the sequence of pilot signals, embedded in the model's latent space, is processed through scaled dot-product attention, producing attention scores that indicate the relative importance of each key to the query.
In Fig. \ref{fig:heatmap1}, \ref{fig:heatmap3}, and \ref{fig:heatmap4}, we visualize the attention scores of the first, third, and fourth heads of the transformer, respectively, when inferring the results of Fig. \ref{fig:beampattern of RTF}. Due to space limitations, we present only the selected attention head scores. As shown, each head exhibits distinct attention patterns, indicating that they capture different contextual representations of the input pilot signal sequence. For instance, when generating polarization and beamforming vectors for the third pilot stage, the first head (Fig. \ref{fig:heatmap1}) assigns $55\%$ relevance to the first pilot signal and $45\%$ to the second. In contrast, the third head (Fig. \ref{fig:heatmap3}) assigns $14\%$ relevance to the first pilot signal and $85\%$ to the second. These differences highlight the ability of multi-head attention to extract diverse features from the received pilot signals, enabling a more robust and adaptive learning process. 

To further analyze the contribution of each attention head, we visualize the array response generated by individual heads. Fig. \ref{fig:beampattern of different attention head} illustrates the array responses of the polarization matrices and beamforming vectors produced by each head. Specifically, the first, second, and third rows of Fig. \ref{fig:beampattern of different attention head} correspond to the array responses generated by Head 1, Head 3, and Head 4, respectively, at pilot rounds $l=0,1,6,9$. As shown in the figure, each head produces a distinct array response based on its independently learned parameters. Notably, individual heads do not exhibit a strong ability to directly align their main lobes with the optimal beamforming direction, as seen in the final column where none of the array responses fully align with the main direction. However, the key advantage of the transformer architecture lies in its ability to integrate diverse contextual representations through the (FFN) layer, ultimately refining the learned beamforming direction as shown by last figure of Fig. \ref{fig:beampattern of RTF}. This capability provides a significant advantage over GRU-based models, which can only capture a single temporal correlation of the input signal. Unlike transformers, which aggregate multiple attention heads to extract diverse features, the GRU can only generate a single summarized representation of the input sequence. As illustrated in Fig. \ref{fig:beampattern of GRU}, this limitation prevents the GRU from effectively identifying the main lobe direction, resulting in suboptimal beamforming performance.

\subsection{Discussion}
In this paper, we have demonstrated that transformers offer significant advantages over RNN architectures for sequential problems. Specifically, we have shown that the multi-head attention mechanism contributes to performance gains by effectively capturing richer contextual information from temporally correlated data. Another key advantage of transformers is their ability to model long-range dependencies more effectively than RNNs. Based on this, we predict that transformers will perform even better in problem settings that require longer time dependencies among the input data. Exploring this aspect is left for future work.

\section{Conclusion}
\label{sec:Conclusion}
Joint optimization of polarization and beamforming vectors in a massive PR-MIMO system with the limited number of RF chains, is a challenging task, in particular, without perfect CSI. The primary difficulty arises from high-dimensional pilots, leading to significant pilot overhead. This paper proposes an adaptive approach that designs and updates polarization and beamforming vectors based on the sequence of received pilot symbols. This is achieved through the proposed transformer-based framework that effectively leverages the multi-head attention mechanism to extract diverse contextual relationships from temporally correlated pilot signals. Numerical results demonstrate that the proposed framework significantly outperforms not only non-adaptive data-driven methods but also RNN-based adaptive approaches. Furthermore, this paper provides interpretation and analysis of the sequential process inside the transformer-based DNN framework, highlighting the contribution of multi-head attention to significant performance gain.
\bibliographystyle{IEEEtran}
\bibliography{references}

\end{document}